\documentclass[reprint,amsmath,amssymb,aps,prl,floatfix,superscriptaddress,nofootinbib,preprintnumbers]{revtex4-2}

\usepackage{graphicx} % Include figure files
\usepackage{dcolumn}  % Align table columns on decimal point
\usepackage{bm}       % Bold math
\usepackage{xcolor}   % For text colors (optional)
\usepackage{orcidlink}
\usepackage{hyperref}
\usepackage{siunitx}
\hypersetup{%
    colorlinks  = true,%
    linkcolor   = blue,%
    citecolor   = blue,%
    urlcolor    = red,%
    linktocpage = true%
}
\usepackage{physics}
\usepackage{comment}

\DeclareSIUnit \parsec {pc}
\DeclareSIUnit \MSolar {\mathrm{M}_\odot}

\begin{document}

%\preprint{APS/123-QED}

\title{\Large Dynamical Heating from Dark Compact Objects and \\ Axion Minihalos: Implications for the 21-cm Signal}% Force line breaks with \\

\author{Badal Bhalla}
\email{badalbhalla@ou.edu}
\affiliation{Homer L. Dodge Department of Physics and Astronomy, University of Oklahoma, Norman, OK 73019, USA}
 
\author{Aurora Ireland}
\email{anireland@stanford.edu}
\affiliation{Leinweber Institute for Theoretical Physics at Stanford, Department of Physics, Stanford University, Stanford, CA 94305, USA}

\author{Hongwan Liu}
 \email{hongwan@bu.edu}
\affiliation{Physics Department, Boston University, Boston, MA 02215, USA}

\author{Huangyu Xiao}
 \email{hxiao3@bu.edu}
\affiliation{Physics Department, Boston University, Boston, MA 02215, USA}
\affiliation{Department of Physics, Harvard University, Cambridge, MA, 02138, USA}
\affiliation{Theoretical Physics Division, Fermi National Accelerator Laboratory, Batavia, IL, USA}

\author{Tao Xu}
 \email{tao.xu@ust.hk}
\affiliation{Homer L. Dodge Department of Physics and Astronomy, University of Oklahoma, Norman, OK 73019, USA}
\affiliation{Jockey Club Institute for Advanced Study, Hong Kong University of Science and Technology, Clear Water Bay, Kowloon, Hong Kong}

\begin{abstract}
The temperature of baryons at the end of the cosmic dark ages can be inferred from observations of the 21-cm hyperfine transition in neutral hydrogen. Any energy injection from the dark sector can therefore be detected through these measurements. Dark compact objects and dark-matter substructures can modify the baryon temperature by transferring heat via dynamical friction. In this work, we evaluate the prospects for detecting dynamical friction–induced heating from dark compact objects with a mass in the range \SIrange{e2}{e5}{\MSolar}, as well as from axion minihalos, using upcoming 21-cm experiments. We find that both the 21-cm global signal and power-spectrum measurements will be sensitive to dark compact objects that constitute about 10\% of the dark matter, and will substantially improve our sensitivity to axion-like particles with masses in the range \SIrange{e-18}{e-9}{\eV}.
\end{abstract}

\maketitle

\textbf{Introduction:} Massive Compact Halo Objects (MACHOs) are a class of dark matter (DM) candidates that lie beyond the conventional particle DM paradigm. Proposed examples include primordial black holes (PBHs)~\cite{Green:2020jor,Carr:2020xqk}, axion stars~\cite{Ruffini:1969qy,Liebling:2012fv,Chang:2024fol}, and compact objects formed through dynamics in the dark sector~\cite{Witten:1984rs,Bai:2018dxf,Kusenko:1997si,Hong:2020est}. Many searches for these objects target unique signatures, such as explosive events from axion stars~\cite{Escudero:2023vgv,Fox:2023xgx,Fox:2025tqa}. 
For PBHs, heating due to the accretion of material around these objects can impact the cosmic microwave background (CMB)~\cite{Ricotti:2007au,Ali-Haimoud:2016} and the properties of dwarf galaxies~\cite{Lu:2020bmd,Takhistov:2021aqx,Kim:2025gck}. Such PBH bounds are, however, not applicable to MACHOs that are less dense, which may not accrete efficiently. 

More generic MACHO constraints come from gravitational microlensing and dynamical heating. Data from the MACHO~\cite{MACHO:2000qbb}, EROS-2~\cite{EROS-2:2006ryy}, and OGLE-IV~\cite{OGLE:2015} microlensing surveys have ruled out MACHOs in the $\lesssim \mathcal{O}(10) \, M_\odot$ mass range as the primary component of DM~\cite{Blaineau:2022nhy}. Meanwhile, higher masses $\gtrsim \mathcal{O}(10) \, M_\odot$ are constrained by the non-observation of dynamical heating of gas and stars in dwarf galaxies~\cite{Brandt:2016aco,Lu:2020bmd,Wadekar:2022ymq,Takhistov:2021aqx,Graham:2023unf,Graham:2024hah,Graham:2025opw,Kim:2025gck}. While microlensing and dynamical heating in small-scale structures offer powerful constraints, each is subject to distinct sources of uncertainty, making complementary approaches valuable. 

In this \textit{Letter}, we propose a novel probe of MACHO abundances based on dynamical friction (DF) heating of the baryon fluid in the early universe. As a MACHO moves through baryons, it generates an enhanced-density wake, and the gravitational interaction between the MACHO and its wake gives rise to a drag force. The resulting momentum loss transfers energy to the baryon fluid, heating it. Because this fluid is collisional, the effect is strongly suppressed at early times while the bulk relative velocity $v_{\rm rel}$ is smaller than the photon-baryon sound speed $c_s$ ($\mathcal{M} \equiv v_{\rm rel} / c_s \ll 1$)~\cite{Ostriker_1999}. Following baryon decoupling and the sharp drop in $c_s$ at recombination, however, typical velocities of $v_{\rm rel} \sim\,$\SI{30}{\kilo\meter\per\second}~\cite{Tseliakhovich:2010bj} correspond to $\mathcal{M} \sim 5$ in the baryon fluid, and so DF becomes efficient. 

The resulting increase in the baryon temperature $T_{\rm B}$ could be detectable with 21-cm cosmology. The heating or cooling of baryons due to momentum transfer between the dark sector and baryons during the cosmic dark ages---and its consequences for the 21-cm global signal and power spectrum---has been extensively studied~\cite{Munoz:2015bca,Munoz:2017qpy,Fialkov:2018xre,Barkana:2018lgd,Munoz:2018pzp,Barkana:2018qrx,Munoz:2018jwq,Liu:2019knx,Barkana:2022hko}. We therefore expect 21-cm observations to be similarly sensitive to DF heating. Compared to the phenomenology of DM-baryon scattering, however, DF heating is remarkable for at least three reasons.
First, the suppression of DF at small $\mathcal{M}$ means that DF is effectively negligible prior to recombination, leading to virtually no impact on the CMB, Lyman-$\alpha$ forest, or small-scale structure. This contrasts sharply with DM-baryon scattering, which affects the evolution of perturbations on scales that enter the horizon before recombination (see \textit{e.g.} Refs.~\cite{Dvorkin:2013cea,Boddy:2018wzy,Slatyer:2018aqg,Xu:2018efh,DES:2020fxi,Buen-Abad:2021mvc,Nguyen:2021cnb}). Second, energy injection from DM-baryon scattering, as well as from DM annihilation and decay, is deposited into several different channels, including ionization, excitation, and heating. Such deposition can then modify both the ionization history and thermal history of the universe.

In contrast, here, all of the deposited energy goes directly into heating. Third, the strong correlation between DF and $v_{\rm rel}$ leads to an \textit{enhancement} in the fluctuations of $T_\mathrm{B}$, and therefore in the 21-cm power spectrum, relative to $\Lambda$CDM expectations.

The dependence of $T_\mathrm{B}$ on $v_{\rm rel}$ also imprints an enhanced baryon acoustic oscillation (BAO) feature onto the power spectrum relative to $\Lambda$CDM~\cite{Dalal:2010yt,Ali-Haimoud:2013hpa,Munoz:2019rhi}. 
This enhancement is also predicted in models with a millicharged DM subcomponent that can transfer heat between baryons and the dark sector~\cite{Barkana:2022hko}, except that here the baryons are heated rather than cooled.

The enhancement stands in contrast to the suppression of the power spectrum generically expected from DM annihilation and decay~\cite{Evoli:2014pva,Lopez-Honorez:2016sur,Facchinetti:2023slb,Sun:2023acy,Sun:2025ksr}, or from PBH accretion~\cite{Agius:2025xbj}, where energy injection leads to a more homogeneous 21-cm signal. 

These striking signatures of DF heating by MACHOs make it an excellent new-physics target for near-future measurements of the 21-cm global signal~\cite{Voytek:2013nua,Philip:2019,deLeraAcedo:2022kiu} and the power spectrum~\cite{HeraII,SKA,SKAlow}. In this work, we focus on two well-motivated scenarios: 1)~MACHOs of a characteristic mass in the \SIrange{e2}{e5}{\MSolar} range comprising some fraction $f_{\rm M}$ of the DM, and 2)~axion (or more generally, axion-like particle (ALP)) minihalos.
%formed from some characteristic minihalo mass at formation, $M_{\rm h,c}$
We estimate that both global signal and power spectrum measurements will be sensitive to $f_{\rm M} \sim 0.1$ for MACHOs in the mass range \SIrange{e4}{e5}{\MSolar}, and will significantly improve sensitivity to the ALP symmetry breaking scale $f_a$ for ALP masses $m_a$ in the range \SIrange{e-18}{e-9}{\eV}. In combination with limits from the apparent absence of black hole superradiance~\cite{Mehta:2020kwu,Baryakhtar:2020gao,Unal:2020jiy,2025MNRAS.tmp.1518H,Witte:2024drg}, these measurements could potentially rule out ALPs with $m_a \sim \SI{e-12}{\eV}$ for any value of $f_a$. 

The DF heating signal depends primarily on the perturber mass and on the well-understood distribution of bulk relative velocities between DM and baryons from linear cosmology. Our estimates are therefore relatively free from the uncertainties that affect other probes like microlensing and UFD heating, such as assumptions about MACHO survivability or accretion, the spatial distribution of MACHOs in galaxies, and baryonic physics. While DF heating has its own uncertainties which will likely require simulations to determine precisely, our work strongly suggests that 21-cm measurements can achieve unprecedented sensitivity to MACHOs and axion minihalos, and strong complementarity to other constraints in regimes where it is expected to be less competitive.\\

\textbf{Dynamical Friction Heating:} For a sufficiently massive DM component streaming through the baryonic fluid, energy exchange will occur through purely gravitational interactions. In particular, the gravitational attraction between the perturber and its enhanced-density wake results in a drag force known as dynamical friction (DF). The energy lost by the perturber in this manner goes into heating the baryonic fluid. 

Refs.~\cite{Just:1986,1990A&A...232..447J,Ostriker_1999} studied DF in a collisional medium, deriving the magnitude of the drag force on a perturber of mass $M$ in the linear regime---where density perturbations due to the perturber are small---as
\begin{equation}\label{eq:DF}
    F_{\rm DF}(M,v_{\rm rel}) = \frac{4 \pi G^2 M^2 \rho_{\rm B}}{v_{\rm rel}^2} \,  \mathcal{I}(\mathcal{M}, \Lambda) \,,
\end{equation}
with $\rho_{\rm B}$ the cosmological mean density of baryons, the medium of interest to us. The function $\mathcal{I}(\mathcal{M},\Lambda)$ depends on the Mach number $\mathcal{M}$ and on the Coulomb logarithm $\ln \Lambda \equiv \ln (b_{\max} / b_{\min})$, which serves to regulate an integral over impact parameters. The minimum and maximum values $b_{\min}$ and $b_{\max}$ can be taken to be the smallest and largest length scales for which the approximations used (point-like perturber, coherent medium response, \textit{etc}.) remain valid. 
An expression for $\mathcal{I}(\mathcal{M},\Lambda)$ valid for all values of $\mathcal{M}$ is provided in the Supplementary Materials (SM). 
Here we simply note that it is maximized for $\mathcal{M} \sim 1$, and has the asymptotic limiting behavior
\begin{alignat}{1}
    \mathcal{I}(\mathcal{M},\Lambda) \approx \begin{cases}
        \mathcal{M}^3 / 3 \,, & \mathcal{M} \ll 1 \,, \\
        \ln \Lambda \,, & \mathcal{M} \gg 1 \,.
    \end{cases}
\end{alignat}
In the subsonic $\mathcal{M} \ll 1$ regime, applicable before recombination when $c_s \sim 1/\sqrt{3}$ and the typical baryon-DM bulk relative velocity is $\sim10^{-5}$~\cite{Dvorkin:2013cea}, we see that $F_{\rm DF}$ is extremely suppressed and therefore negligible. Physically, the large sound speed results in an almost front-back symmetric density distribution of baryons about the perturber. Following recombination, however, the baryon sound speed drops to $c_s \sim \sqrt{T_{\rm B}/m_{\rm p}} \sim \SI{10}{\kilo\meter\per\second}$. Since typical bulk relative velocities at this time are $v_{\rm rel} \sim \SI{30}{\kilo\meter\per\second}$~\cite{Tseliakhovich:2010bj}, this corresponds to the regime $\mathcal{M} \gtrsim 1$, for which DF becomes relevant. 
Physically, the supersonic regime is characterized by an enhanced-density region confined to a Mach cone behind the perturber with a half-opening angle $\sin \theta = 1/\mathcal{M}$. Larger $\mathcal{M}$ then corresponds to a narrower Mach cone and a more geometrically asymmetric wake.

In this regime, the value of $\Lambda$, set by a choice of $b_{\rm min}$ and $b_{\rm max}$, is important for determining the DF force.  
We take $b_{\rm max} = \mathcal{M} \lambda_{\rm J}$, where $\lambda_{\rm J} = \sqrt{\pi c_s^2 / G \rho_{\rm B}} \sim c_s H^{-1}$ is the Jeans length, with $H$ the Hubble parameter. This is roughly the distance traveled by the perturber over the age of the Universe, a reasonable estimate for the transverse size of the Mach cone and the effective size of the surrounding gaseous medium, and is consistent with Refs.~\cite{McQuinn:2012,OLeary:2012}. For $b_{\min}$, we take the Bondi-Hoyle-Lyttleton radius for MACHOs, $b_{\rm min} = G M/(c_s^2 + v_{\rm rel}^2)$, based on the conclusion drawn by Ref.~\cite{Suzuguchi:2024btk} that there is no DF from gas within this radius.
This is therefore a good choice for MACHOs that have a smaller spatial extent than the Bondi-Hoyle-Lyttleton radius. 
For axion minihalos, we adopt $b_{\rm min} = 0.35 \mathcal{M}^{0.6} r_{\rm vir}$, where $r_{\rm vir}$ is the virial radius of the halo, consistent with the choice made in the simulations of Refs.~\cite{Kim_2009,McQuinn:2012,OLeary:2012} to match the linear results of Refs.~\cite{Just:1986,1990A&A...232..447J,Ostriker_1999}.  

The DF heating rate per baryon $\dot{Q}_{\rm DF}$ is~\cite{Conroy_2008,Kim_2005} 
\begin{alignat}{1}\label{eq:Qdf}
    \dot{Q}_{\rm DF}(v_{\rm rel},z) = \frac{1}{n_{\rm B}} \int {\rm d}M \frac{\rho_{\rm DM}}{M} \frac{{\rm d} f}{{\rm d} M} v_{\rm rel} F_{\rm DF} \,,
\end{alignat}
where $v_{\rm rel} F_{\rm DF}$ is the perturber's rate of kinetic energy loss, $\rho_{\rm DM}$ is the DM energy density, and ${\rm d} f / {\rm d} M$ is the fraction of DM (by mass) in perturbers of mass $M$. 
The resulting change in baryon temperature can be estimated by numerically solving the following equations:
\begin{align}
    \frac{{\rm d}v_{\rm rel}}{{\rm d}\ln a} & = - v_{\rm rel} - \frac{F_{\rm DF}}{MH} \,, \label{eq:drag} \\
    \frac{{\rm d} T_{\rm B}}{{\rm d}\ln a} & = -2T_{\rm B} + \frac{\Gamma_C}{H}(T_{\rm CMB} - T_{\rm B}) + \frac{2}{3} \frac{\dot{Q}_{\rm DF}}{H} \,,
    \label{eq:TB}
\end{align}
where $T_{\rm CMB}$ is the CMB radiation temperature and $\Gamma_C$ is a rate governing Compton scattering. 
We also simultaneously evolve the free electron fraction using the three-level atom model~\cite{1968ApJ...153....1P,1968ZhETF..55..278Z}; more details can be found in Ref.~\cite{Liu:2019bbm}. 
The terms in Eq.~\eqref{eq:TB} correspond only to adiabatic cooling, heating through Compton scattering with the CMB, and DF heating, respectively. In this work, we only consider the 21-cm signal for $z \geq 17$, and assume no significant X-ray heating has occurred until this point, which is a reasonable scenario given current astrophysical constraints (see \textit{e.g.}~Ref.~\cite{Reis:2021nqf}).  
We also neglect other irreducible sources of heating, including Lyman-$\alpha$ recoils~\cite{Chen:2003gc,Chuzhoy:2005wv,Furlanetto:2006fs,Venumadhav:2018uwn}, since these are subdominant to the $\mathcal{O}(1)$ heating contribution from DF we will be sensitive to.

These equations are integrated from recombination at $z = 1100$, starting from an initial bulk relative velocity $\vec{v}_{\rm rel}^{\,*}$, down to $z = 17$. 
Since the DF effect is negligible prior to recombination, the velocity distribution at recombination is Gaussian, as predicted by $\Lambda$CDM, \textit{i.e.}\ 
\begin{equation}
    f(\vec{v}_{\rm rel}^{\,*}) = \left( \frac{3}{2 \pi v_{\rm rms}^2} \right)^{3/2}\text{exp}\left(- \frac{3v_{\rm rel}^{*2}}{2v_{\rm rms}^2} \right) \,,
    \label{eq:f_vrel}
\end{equation}
where we set $v_{\rm rms} = \SI{29}{\kilo\meter\per\second}$~\cite{Tseliakhovich:2010bj}. 
We thereby obtain $T_{\rm B}(v_{\rm rel}^*, z)$ for $0 \leq v_{\rm rel}^* \leq \SI{100}{\kilo\meter\per\second}$, which encompasses most of the range relevant for an accurate 21-cm determination. 

Note that under our approximations, $T_{\rm B}$ depends only on $v_{\rm rel}^*$ and $z$; properly capturing other effects would require simulations, which we leave to future work. 

Although the preceding results were derived assuming that all density perturbations are small, for $\mathcal{M} \gtrsim 1$, this is not a self-consistent assumption; in fact, near the edges of the Mach cone, the fluid overdensity becomes large, approaching infinity for a point-like perturber. 
Despite this inconsistency, Ref.~\cite{Kim_2009} showed that there is a tight correlation between the DF force estimated in simulations and $F_{\rm DF}$ calculated in the linear regime using Eq.~\eqref{eq:DF}, finding a large, order of magnitude difference only for very massive halos, extremely small perturbers, or when $\mathcal{M}$ is very close to 1. 
We find that this discrepancy should not affect our MACHO analysis significantly, but could lead to an $\mathcal{O}(1)$ overestimate of the DF force for heavy axion minihalos in a limited redshift window; see the SM for further details.\\

\textbf{MACHOs \& Axion Minihalos:} We model dark compact object DF heating in two scenarios. First, we adopt a phenomenological approach and simply assume a population of MACHOs with a single characteristic mass $M$ making up a fraction $f_{\rm M}$ of the DM energy density, \textit{i.e.}~$\dd f / \dd M' = f_{\rm M} \delta(M'-M)$. 
Second, we consider DF heating from axion minihalos; this scenario has a complicated perturber mass function that evolves with redshift, depending on the ALP parameters $m_a$ and $f_a$.

\begin{figure*}[t]
  \centering
  \includegraphics[width=0.32\linewidth]{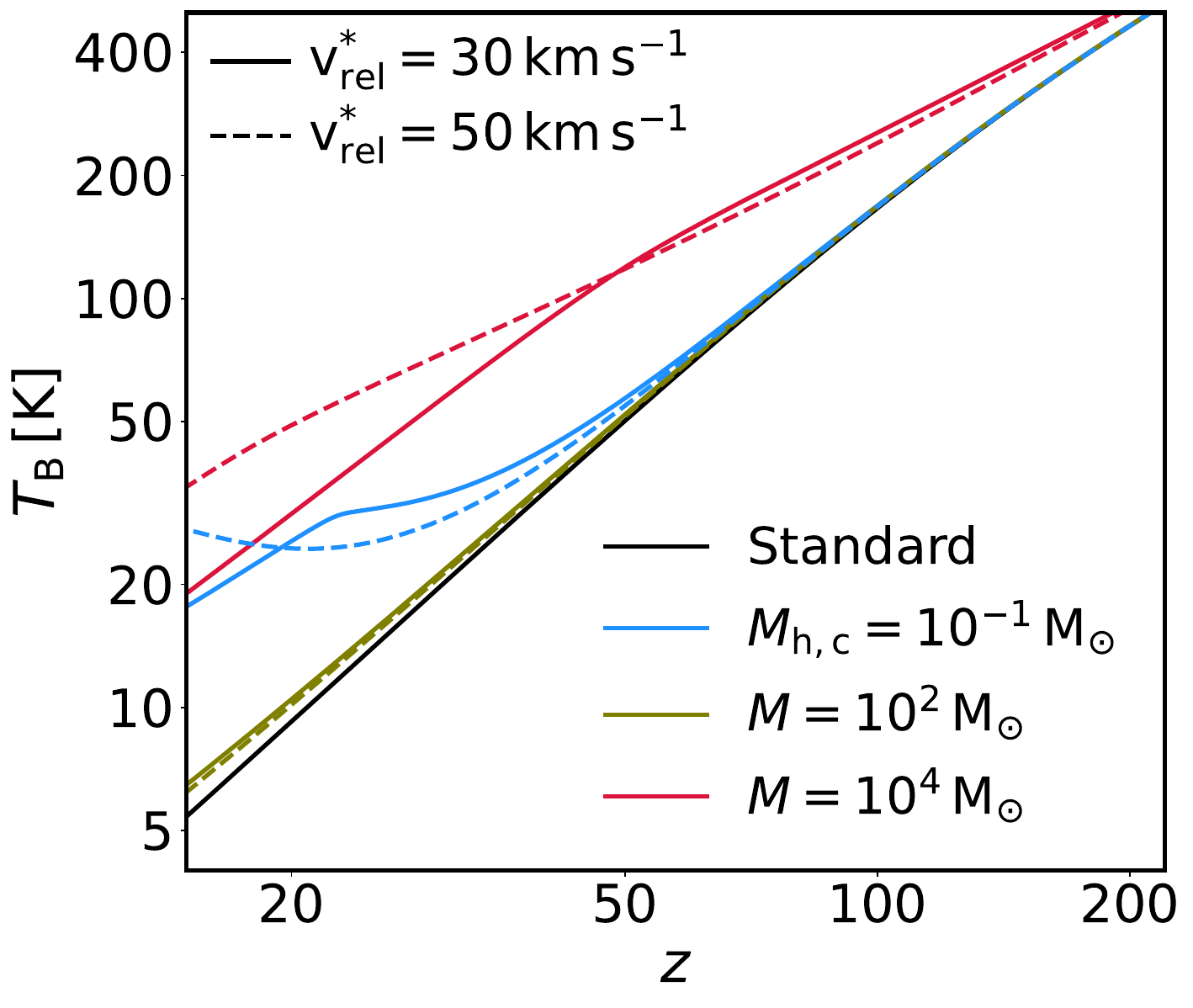}\hfill
  \includegraphics[width=0.32\linewidth]{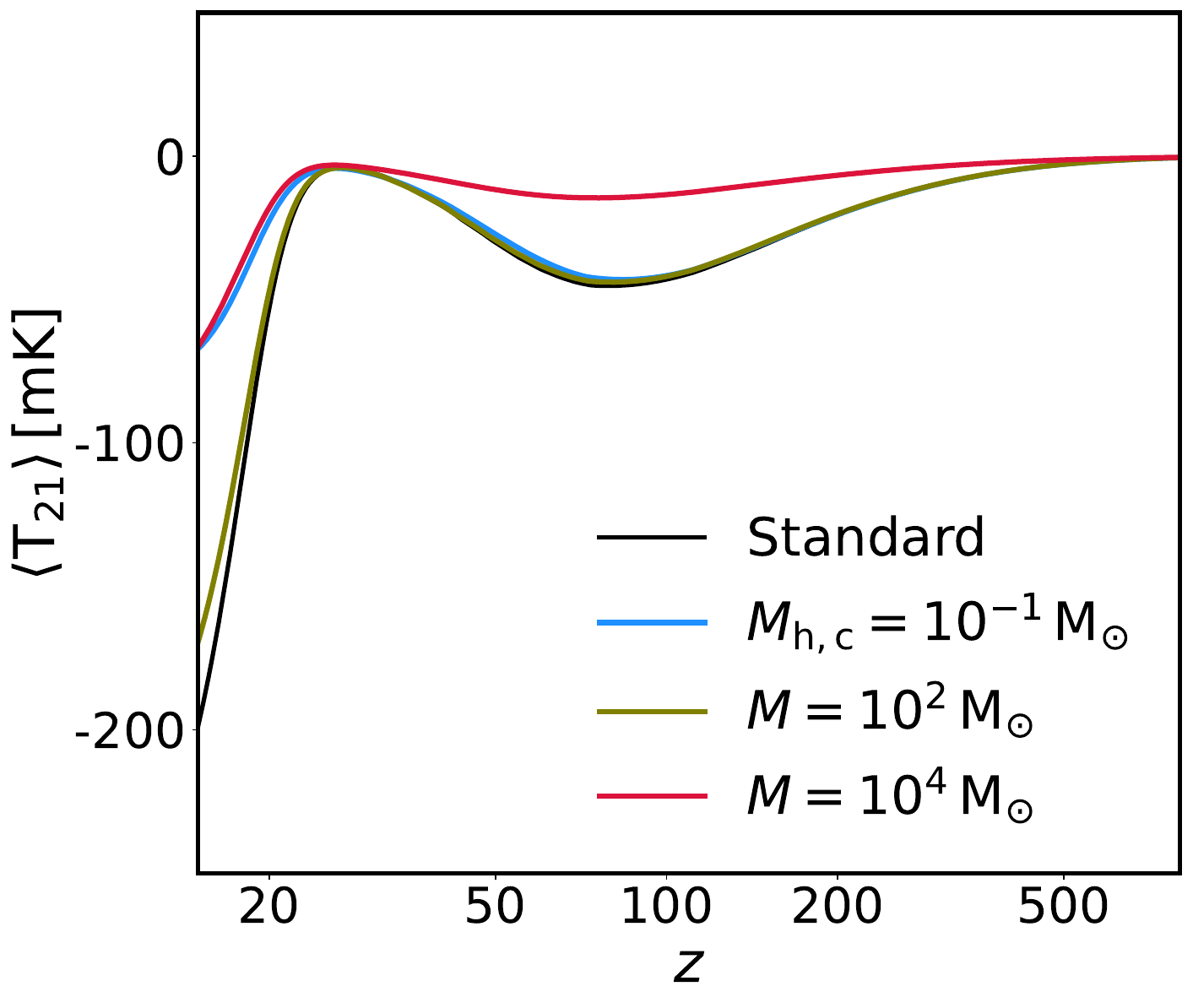}\hfill
  \includegraphics[width=0.335\linewidth]{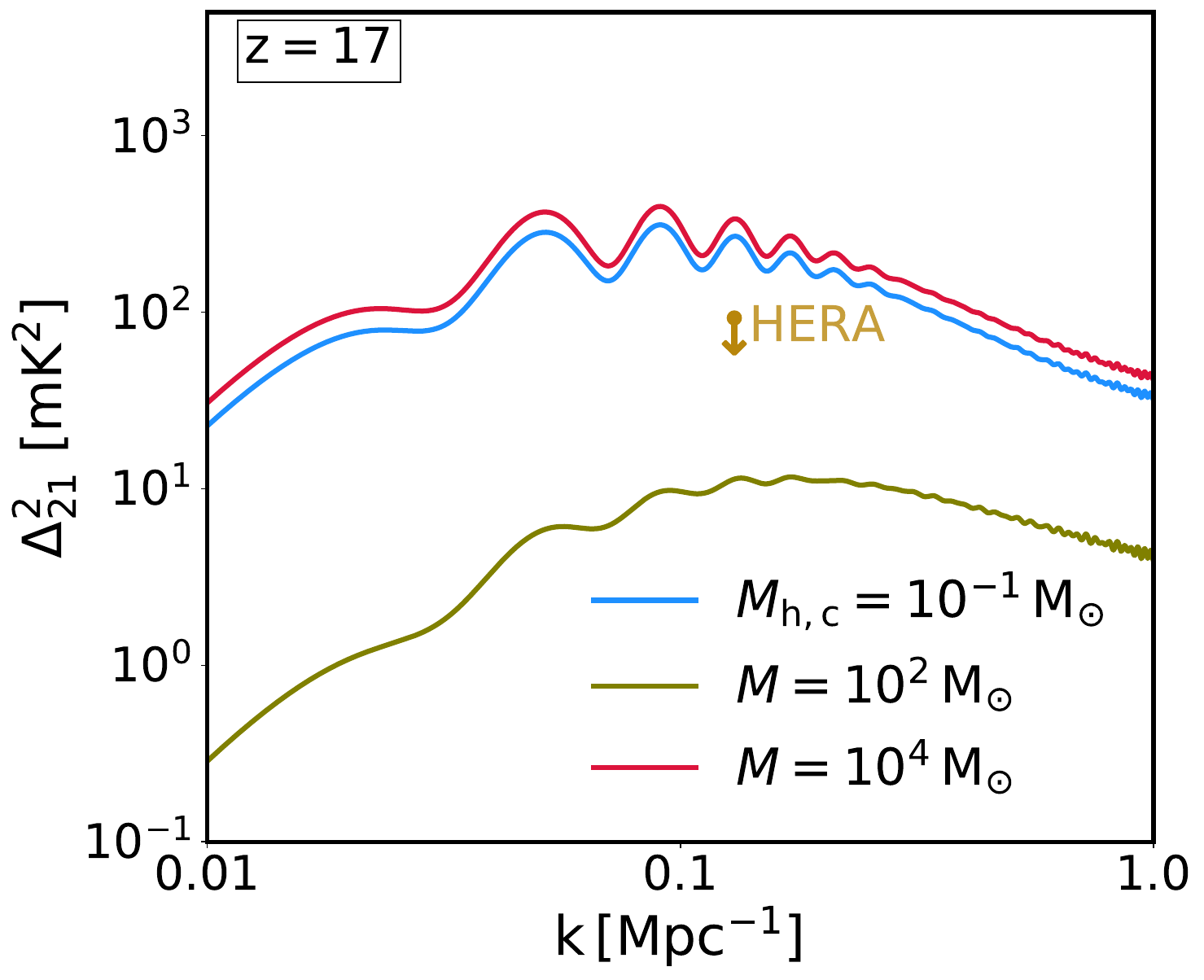}
  \caption{\textbf{Left:} Baryon temperature $T_{\rm B}$ as a function of redshift for MACHOs of mass $M = 10^2 \, \rm M_{\odot}$ (olive) and $M = 10^4 \, \rm M_{\odot}$ (red), and for axion minihalos with an initial mass $M_{\rm h,c} = 10^{-1} \, \rm M_{\odot}$ (blue), assuming MACHOs/axions constitute 100\% of the DM. The solid (dashed) lines correspond to a DM-baryon relative velocity at recombination of $v_{\rm rel}^* =30 \, (50) \SI{}{\kilo\meter\per\second}$. \textbf{Middle:} The sky-averaged 21-cm brightness temperature $\langle T_{21} \rangle(z)$. \textbf{Right:} The 21-cm power spectrum $\Delta_{21}^2$ as a function of comoving wavenumber at $z = 17$. The golden arrow indicates the expected HERA sensitivity of \SI{92.7}{\milli\kelvin\squared} after $10^3$ hours of integration time, following Refs.~\cite{Munoz:2018jwq,Barkana:2022hko}.}
  \label{fig:21}
\end{figure*}

Axion minihalos (or miniclusters) will form as early as matter-radiation equality due to the enhanced small-scale fluctuations which are produced by the randomized axion field across different horizons in the post-inflationary scenario ~\cite{Hogan:1988mp,Kolb:1993zz,Kolb:1995bu}. 
Even in scenarios with pre-inflationary Peccei-Quinn breaking, many production mechanisms can lead to the formation of these dense axion substructures~\cite{Hardy:2016mns,Fukunaga:2020mvq,Arvanitaki:2019rax,Co:2019jts,Eroncel:2022vjg,Eroncel:2022efc}. The initial axion minihalo mass is determined by the DM mass contained within the horizon at the axion oscillation time, when axions start to behave as nonrelativistic matter. Since the initial axion density is $\sim m_a^2 f_a^2$, the oscillation time can be uniquely determined from the relic abundance and axion parameters. Therefore, the initial axion minihalo mass is~\cite{Fox:2025tqa}
\begin{equation}\label{eq:minihalo_mass}
    M_{\text{h},\text{c}}\!\approx\! \SI{1.5e-11}{\MSolar}\!\left( \frac{m_a}{\SI{e-6}{\eV}} \right)^{-2}\!\!\left(\!\frac{f_a}{\SI{e11}{\giga\eV}}\!\right)^{-2}\!\!\!\!{g_{*,\rm osc}^{-1/2}},
\end{equation}
where $g_{*,\rm osc}$ is the degrees of freedom at the onset of oscillation. Although the value of $g_{*,\rm osc}$ depends on $m_a$ and $f_a$, we conservatively take it to be 10, which is also consistent with the relevant axion parameters where the bound is the strongest. To model the axion minihalo mass function, we use a semi-analytic Press-Schechter model~\cite{1974ApJ...187..425P}, ${{\rm d}f_{\rm h}}/{{\rm d} \ln M_{\rm h}} = \sqrt{{\nu}/{(2\pi)}}e^{-\nu/2} \, {{\rm d} \ln \nu}/{{\rm d} \ln M_{\rm h}} \,,
$
where $f_{\rm h}$ is the fraction of DM bound in minihalos, $M_{\rm h}$ is the minihalo mass, $\nu \equiv \delta_c^2/[\sigma^2(M_{\rm h})D^2(z)]$, $\delta_c = 1.68$ and $D(z)$ is the growth function. 
 
This mass function agrees reasonably well with other analytic and numerical studies on the axion minihalo mass function \cite{Enander:2017ogx,Eggemeier:2019khm,Ellis:2020gtq,Xiao:2021nkb}. 
Since fluctuations on small scales are dominated by the order unity axion perturbations, one can consider the axion-induced power spectrum independently from the adiabatic fluctuations. In the post-inflationary scenario, a totally randomized axion field is converted to DM inhomogeneities. Therefore, it is well described by a white-noise spectrum, with the characteristic scale determined by the horizon size at the oscillation time. Given the spectrum, the variance entering in the minihalo mass function is computed as $\sigma(M_{\rm h})\approx \sqrt{3\pi M_{\rm h,c}/(2M_{\rm h})}$.

The white-noise power spectrum accurately characterizes the initial minihalo formation and evolution until massive cold dark matter (CDM) halos form from the adiabatic fluctuations around $z \sim 20$. After that, some fraction of the axion minihalos will be subsumed into the massive CDM halos, effectively freezing their population. The final axion minihalo mass function, including the infall minihalos, is heuristically~\cite{Xiao:2021nkb} 
\begin{multline}
	%\frac{M}{\rho_{\rm DM}} \frac{\dd n_{\rm M}}{\dd M} =
    \frac{{\rm d}f}{{\rm d}M_{\rm h}}(z) =\int_{z_{\rm eq}}^{z} \!\!{\rm d}z' \frac{{\rm d}f_{\rm col}^{\rm CDM}}{{\rm d}z'}\frac{{\rm d}f_{\rm h}}{{\rm d}M_{\rm h}}(z') \\
    + [1-f_{\rm col}^{\rm CDM}(z)] \frac{\dd f_{\rm h}}{\dd M_{\rm h}}(z) \,,
    \label{eq:axMF}
\end{multline}
where $f_{\rm col}^{\rm CDM}(z)$ is the collapse fraction of massive CDM halos predicted by the adiabatic fluctuations,
\begin{equation}
	f_{\rm col}^{\rm CDM}(z)={\rm erfc}\left(\frac{\delta_{\rm c}}{\sqrt{2}\,\sigma_{\rm CDM}(M_{\rm cut})\,D(z)}\right) \,,
	\label{eqn:fcol}
\end{equation}
where $\sigma^{2}_{\rm CDM}(M)$ is the variance of adiabatic fluctuations and $M_{\rm cut}$ is the minimal mass of CDM halos that can absorb axion minihalos in the parameter space of relevance. We take $M_{\rm cut}=10^6M_{\odot}$ so that the CDM halos are always larger than the axion minihalo for the infall process to be valid, for which $\sigma_{\rm CDM}(M_{\rm cut})=8.2$ \cite{Murray:2013qza}. Note that the collapse fraction depends nearly logarithmically on the choice of $M_{\rm cut}$ due to the power spectrum of adiabatic fluctuations. The collapse fraction of massive CDM halos is only $\sim 1\%$ at $z\sim 17$.  Therefore, the majority of axion minihalos remain isolated, and the contribution from CDM halos will be subdominant, which is consistent with previous studies on the dynamical heating from CDM halos \cite{OLeary:2012,McQuinn:2012}. 

The DF heating rate due to axion minihalos can be evaluated by substituting Eq.~\eqref{eq:axMF} into Eq.~(\ref{eq:Qdf}). We take the lower limit of integration to be $M_{\rm h,c}$ and the upper limit to be $M_{\rm max} = 10^6 \, M_\odot$, which is less than $M_{\rm cut}$, and simultaneously avoids the region of parameter space where the linear approximation is poor (see the SM for more details). 
In any case, halos with such a large mass deposit most of their energy through DF while baryons are still tightly coupled to the CMB, leading to insignificant heating.

In the left panel of Fig.~\ref{fig:21}, we show the evolution of $T_{\rm B}$ as a function of $z$ and $v_{\rm rel}^*$.  
The DF heating is most noticeable at low redshifts, once the baryons have thermally decoupled from the CMB. The heating rate increases with the perturber mass, but decreases with an increase in relative velocity, which is consistent with the drag force scaling as $F_{\rm DF} \propto M^2 \ln (v_{\rm rel}^2)/v_{\rm rel}^2$. 
Note that baryon heating is only efficient after thermal decoupling from the CMB at $z \sim 150$. 
As such, for sufficiently small values of $v_{\rm rel}^*$, the perturber could become subsonic before decoupling, leading to reduced heating, especially for larger perturbers. Conversely, for sufficiently large $v_{\rm rel}^*$, DF-induced heating is suppressed at high redshifts, but becomes significant at low redshifts, after the relative velocity has decreased somewhat.\\

\begin{figure*}[t]
  \centering
  \includegraphics[width=0.455\linewidth]{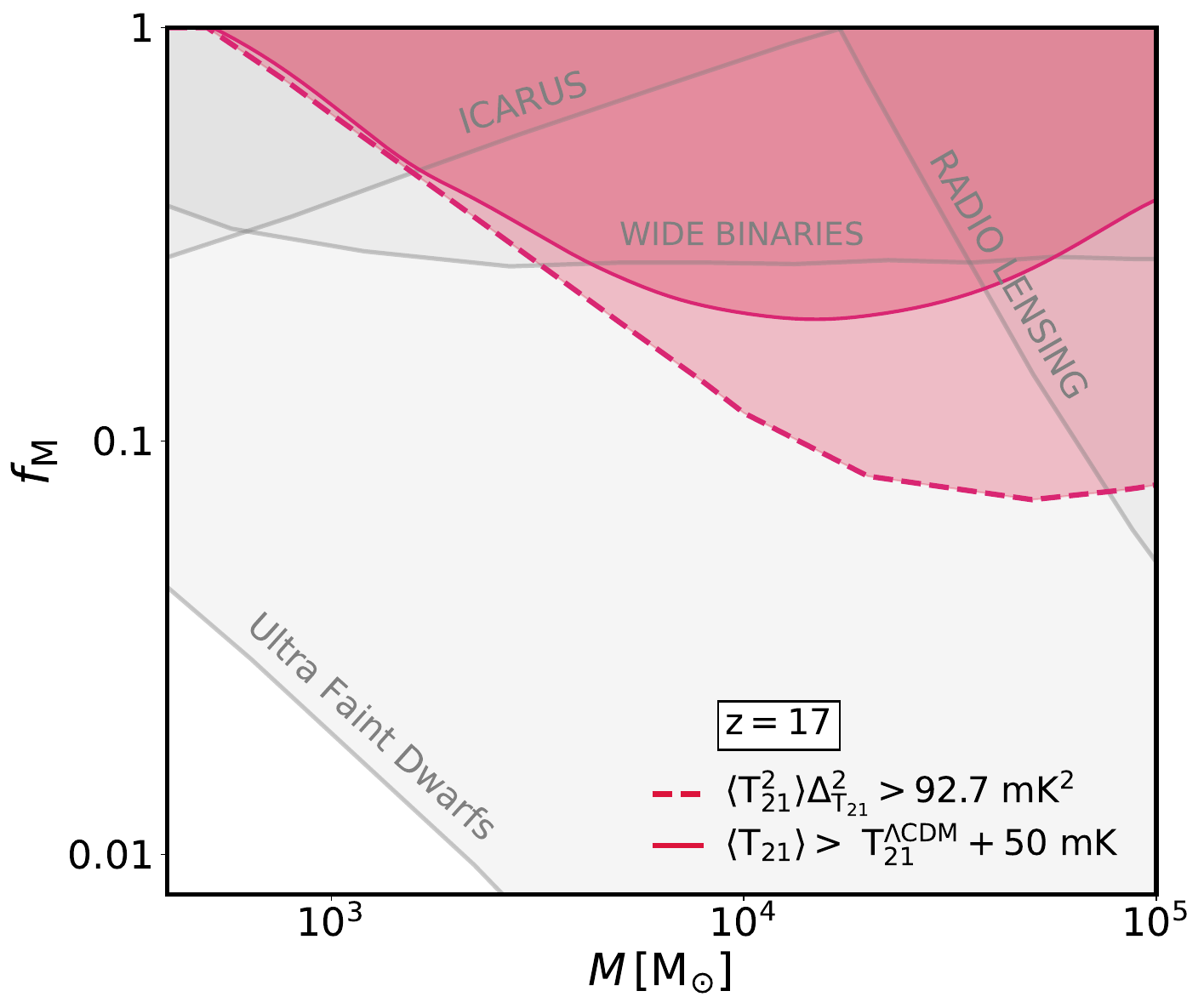}
  \includegraphics[width=0.45\linewidth]{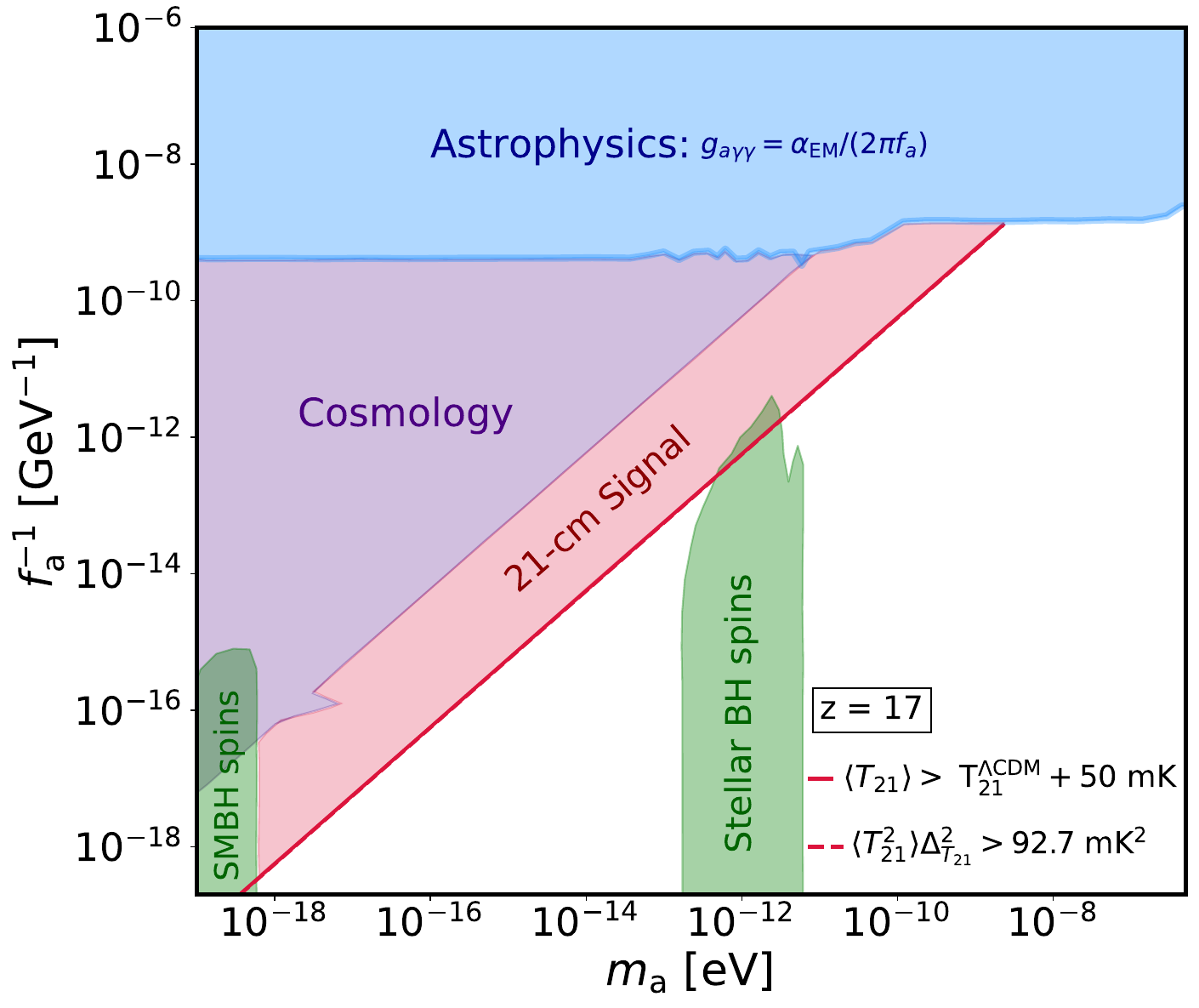}
  \caption{\textbf{Left:} Forecasted sensitivity (in red) of the 21-cm signal to the MACHO fraction as a function of MACHO mass. The solid curve marks the region in which DF heating yields a brightness temperature that is 50 mK higher than expected in $\Lambda$CDM, while the dashed line corresponds to the forecasted sensitivity from HERA. The other limits come from the observed half-light radius of Ultrafaint dwarf galaxies (UFDs)~\cite{Graham:2023unf}, the existence of wide-binaries~\cite{Ramirez:2022mys}, non-detection of lensing effects of compact radio sources~\cite{Zhou:2021tvp}, and the caustic crossing of Icarus~\cite{Oguri:2017ock}. See also Refs.~\cite{Wadekar:2022ymq,Kim:2025gck} for bounds under different assumptions of the MACHO density profile. \textbf{Right:}  Forecasted sensitivity (in red) of the 21-cm signal to ALP symmetry breaking scale $f_a$ for ALP masses $m_a$. The solid and dashed curves correspond to an axion minihalo with an initial mass of $M_{\rm h,c} = 0.015 \, \rm M_{\odot}$. Also shown are bounds derived from astrophysical probes of the axion-photon coupling assuming $g_{a\gamma\gamma} = \alpha_{\rm EM} / (2 \pi f_a)$~\cite{Reynes:2021bpe,Ning:2024eky,Marsh:2017yvc,Muller:2023vjm,Fermi-LAT:2016nkz,Reynolds:2019uqt}, cosmology (in purple)~\cite{Harigaya:2025pox,Murgia:2019duy}, and the superradiance bounds (in green) taken from Refs.~\cite{Mehta:2020kwu,Baryakhtar:2020gao,Unal:2020jiy,2025MNRAS.tmp.1518H,Witte:2024drg}, as compiled in Ref.~\cite{AxionLimits}.}
  \label{fig:sensitivity}
\end{figure*}

\textbf{21-cm Signal:}
The absorption or emission of 21-cm blackbody CMB photons via the hyperfine transition of neutral hydrogen atoms at redshift $z$ leads to a spectral distortion of the CMB at wavelength $\SI{21}{\centi\meter} \times (1 + z)$, typically reported as a differential brightness temperature relative to the CMB along some line-of-sight $\hat{n}$ and redshift $z$, $T_{21}(\hat{n},z)$.
The size of this distortion depends on the relative occupation number of the two-level hyperfine system, parametrized by a single temperature $T_{\rm s}$, the spin temperature, which is strongly dependent on the baryon temperature $T_{\rm B}$.
Detecting the 21-cm signal would therefore provide us with a relatively direct probe of $T_{\rm B}$ during the cosmic dark ages, allowing us to look for anomalous heating due to DF of compact objects. 
There are two main experimental modes for detecting this signal: the sky-averaged or global signal integrated over solid angle $\langle T_{21} \rangle(z)$, or a measurement of the power spectrum of fluctuations $\Delta_{21}^2(k,z)$, where $k$ is the comoving wavenumber. 

The brightness temperature is given by

\begin{equation}\label{eq:T_21_definition}
    T_{21} = \frac{1}{1+z}(T_{\rm s} - T_{\rm CMB})(1-e^{-\tau}) \,,
\end{equation}
where $\tau$ is the optical depth for resonant $21$-cm absorption~\cite{Barkana:2016nyr}. 

The spin temperature is determined by the interactions between the two-level system and CMB photons, other hydrogen atoms, and Lyman-$\alpha$ radiation, according to~\cite{Barkana:2016nyr}
\begin{equation}
    T_{\rm s}^{-1} = \frac{T_{\rm CMB}^{-1} + x_\alpha T_\alpha^{-1} + x_{c}T_{\rm B}^{-1}}{1 + x_\alpha + x_c} \,,
\end{equation}
where $T_\alpha$ is the color temperature of the Ly$\alpha$ radiation field, and $x_\alpha$, $x_c$ are the Ly$\alpha$ and collisional coupling coefficients, respectively.
In our analysis, we set $T_\alpha = T_{\rm B}$ which is an excellent approximation for $T_{\rm B} \gtrsim \SI{1}{K}$~\cite{Barkana:2016nyr}, and use the default values of $x_\alpha(z)$ used in \texttt{Zeus21}~\cite{Munoz:2023kkg}; see Ref.~\cite{Barkana:2022hko} for details on how $x_c(z)$ is computed, based on results in Refs.~\cite{2005ApJ...622.1356Z,Furlanetto:2006su,Furlanetto:2007te}. 

We estimate the global signal and power spectrum at $z = 17$ in order to determine the sensitivity to DF heating by compact objects, using $T_{\rm B}(v_{\rm rel}^*, z)$ determined from integrating Eq.~\eqref{eq:TB} to obtain $T_{21}(v_{\rm rel}^*, z)$. The sky-averaged brightness temperature is then given by
\begin{equation}
    \langle T_{21} \rangle (z) = \int {\rm d}^3 \vec{v}_{\rm rel}^{\,*} \, f(\vec{v}_{\rm rel}^{\,*}) T_{21}(v_{\rm rel}^*,z) \,, 
\end{equation}
where $f(\vec{v}_{\rm rel}^{\,*})$ is given in Eq.~\eqref{eq:f_vrel}. 

For the power spectrum, we define the 21-cm brightness temperature fluctuation as $\delta_{T_{21}}(\vec{x},z) \equiv T_{21}(\vec{x}\,) / \langle T_{21} \rangle(z) - 1$.
Assuming homogeneity and isotropy, the two-point correlation function (2PCF) reads $\xi(r) \equiv \langle \delta_{T_{21}} (\vec{x}_1) \delta_{T_{21}} (\vec{x}_2) \rangle_r$, where $\langle \cdots \rangle_r$ denotes an average over all pairs of points such that $|\vec{x}_1 - \vec{x}_2| = r$. The 21-cm power spectrum is then obtained via Fourier transform, 
\begin{alignat}{1}
    \Delta_{21}^2(k, z) = \frac{2 k^3}{\pi} \langle T_{21} \rangle^2 (z) \int_0^\infty {\rm d}r \, r^2 \xi(r) \frac{\sin(kr)}{kr} \,. 
\end{alignat}

$\xi(r)$ is determined by averaging over the joint Gaussian distribution of the relative velocities at separated points,
\begin{multline}
    \xi(\vec{r}) \!=\!\! \int\! {\rm d}^3 \vec{v}_{{\rm rel},a}^{\,*}\! \int\! {\rm d}^3 \vec{v}_{{\rm rel},b}^{\,*} \mathcal{P}(\vec{v}_{{\rm rel},a}^{\,*}, \vec{v}_{{\rm rel},b}^{\,*}; \vec{r}) \\
    \times \delta_{T_{21}}\!(\vec{v}_{{\rm rel},a}^{\,*}) \delta_{T_{21}} \!(\vec{v}_{{\rm rel},b}^{\,*}) \,,
\end{multline}
 
where the exact structure of $\mathcal{P}$ follows from linear theory~\cite{Dalal:2010yt,Ali-Haimoud:2013hpa,Barkana:2022hko}.

The middle panel of Fig.~\ref{fig:21} shows the global signal, including only DF heating for both cases considered here.
DF heating suppresses the global signal, since $T_{\rm B}$ is brought closer to $T_{\rm CMB}$, leading to less net absorption. 
Note that although the signals for MACHOs with mass $M = \SI{e4}{\MSolar}$ and axion minihalos with $M_{h,c} = \SI{e-1}{\MSolar}$ are similar at $z \lesssim 30$, they differ significantly at $z \sim 100$, since the masses of axion minihalos grow with time, leading to much less heating at early times. 
The corresponding power spectrum is shown in the right panel of Fig.~\ref{fig:21}. 
We show also an approximate forecast of the sensitivity of HERA to the power spectrum after $10^3$ hours of integration time, \textit{i.e.}\ $\Delta_{21}^2 = \SI{92.7}{\milli\kelvin\squared}$ for $k = \SI{0.13}{\per\mega\parsec}$ at $z = 17$~\cite{Munoz:2018jwq,Barkana:2022hko}. This is comparable to the largest magnitude of the power spectrum anticipated in $\Lambda$CDM astrophysical models~\cite{Reis:2021nqf}. 
In the representative cases considered here, the DF signal can significantly exceed this. 

The strong dependence of DF heating on the bulk relative velocity has two effects on $\Delta_{21}^2$. First, it enhances fluctuations in $T_{21}$ significantly, leading to a larger overall amplitude in $\Delta_{21}^2$. 
Second, the baryon acoustic oscillation feature in the bulk relative velocity power spectrum gets imprinted onto $\Delta_{21}^2$. 
The signature is striking; for sufficiently large DF heating, $\Delta_{21}^2$ would be larger than $\Lambda$CDM expectations, making 21-cm fluctuations easier to detect.\\

\textbf{Results:} We estimate the sensitivity of the global signal by requiring that DF heating increases $\langle T_{21} \rangle$ by \SI{50}{\milli\kelvin} relative to the $\Lambda$CDM expectation at $z = 17$. 
For the power spectrum, we instead require that $\Delta_{21}^2$ at $k = \SI{0.13}{\per\mega\parsec}$ and $z = 17$ exceed the HERA forecasted sensitivity discussed above. 
The corresponding projected sensitivities are depicted as dashed and solid lines, respectively, in Fig.~\ref{fig:sensitivity}. 
We find that the 21-cm signal can be sensitive to MACHOs comprising $10\%$ of the DM in the mass range \SIrange{e4}{e5}{\MSolar}. 
This improves on searches using wide-binaries~\cite{Ramirez:2022mys}, lensing of compact radio sources~\cite{Zhou:2021tvp} and the disruption of caustics~\cite{Oguri:2017ock}, but is still outperformed by dynamical heating of stars in UFDs~\cite{Brandt:2016aco,Graham:2023unf}. 

For axion minihalos, the mass function and the behavior of such objects within other systems complicate the ability of lensing or dynamical heating to constrain them. 
The relatively homogeneous conditions of the early Universe, however, means that we have no such concerns. 
We find that the smallest initial halo mass $M_{h,c}$ to which the 21-cm signal is sensitive to is $0.015 \, \rm M_{\odot}$, which grows to $\mathcal{O}(10^3) \, \rm M_{\odot}$ by $z=17$. The corresponding sensitivity to $f_a$ as a function of $m_a$ is shown by the solid red curve in Fig.~\ref{fig:sensitivity}, extending into new parameter space for $\SI{e-18}{\eV} \lesssim m_a \lesssim \SI{e-9}{\eV}$. Together with bounds on the axion–photon coupling in astrophysical environments~\cite{Reynes:2021bpe,Ning:2024eky,Marsh:2017yvc,Muller:2023vjm,Fermi-LAT:2016nkz,Reynolds:2019uqt,Escudero:2023vgv,AxionLimits}, cosmological constraints from the Lyman-alpha forest~\cite{Murgia:2019duy} and free-streaming effects~\cite{Harigaya:2025pox}, as well as bounds from the absence of black hole superradiance~\cite{Witte:2024drg}, the 21-cm signal can completely rule out ALPs with a mass $m_a \sim 10^{-12} \, \text{eV}$.
Note that the limits on $f_a$ derived from the axion-photon coupling $g_{a\gamma\gamma}$ assume the standard proportionality factor of $g_{a\gamma\gamma} = \alpha_{\rm EM} / (2 \pi f_a)$, while all other probes considered here are directly sensitive to $f_a$. 

Although we have only discussed the 21-cm signal just before the onset of cosmic dawn at $z \sim 17$, DF heating from dark substructures will also lead to a 21-cm signal deep in the cosmic dark ages ($z \sim 80$), and provides a strongly motivated new-physics benchmark for plans to build radio telescopes on the far side of the moon~\cite{Burns:2021ndk,2025arXiv250403418B}. 

The estimates here entirely rely on analytical results---backed by numerical simulations---of a single perturber at constant Mach number moving through gas at a fixed sound speed.
The real picture is, however, more complicated: we have instead some number density of perturbers progressively decelerating through baryonic gas that is cooling as the Universe expands, with many wakes dissipating into heat throughout the cosmic volume.
We have also not explored how the signal would change depending on the details of cosmic dawn. 
Finally, since DF changes the bulk relative velocity between DM and baryons at cosmic dawn, this will also impact how baryons fall into DM potential wells when the first structures were formed~\cite{Tseliakhovich:2010bj}---we have not attempted to model the impact of this on the 21-cm signal in any way.
A full treatment of DF and its impact on the 21-cm signal requires dedicated simulations, which we leave for future work.\\

\textbf{Acknowledgments:}
The authors would like to express their gratitude to Dominic Agius, Ethan Baker, Rouven Essig, Peter Graham, Junwu Huang, Marc Kamionkowski, Andrew J. Long,  Matt McQuinn, and Kuver Sinha for insightful discussions. BB is supported by the Dodge and Avenir student fellowships. AI is supported by NSF Grant PHY-2310429, Simons Investigator Award No.~824870, DOE HEP QuantISED award \#100495, the Gordon and Betty Moore Foundation Grant GBMF7946, and the U.S.~Department of Energy (DOE), Office of Science, National Quantum Information Science Research Centers, Superconducting Quantum Materials and Systems Center (SQMS) under contract No.~DEAC02-07CH11359. HL and HX are supported by the U.S. Department of Energy under grant DE-SC0026297. In addition, HL is supported by the Cecile K. Dalton Career Development Professorship, endowed by Boston University trustee Nathaniel Dalton and Amy Gottleib Dalton. TX is supported by the U.S. National Science Foundation under award PHY-2412671. TX also acknowledges support from the Jockey Club Institute for Advanced Study at The Hong Kong University of Science and Technology.

\bibliography{ref}

@string{apj = "ApJ"}

@string{pasp = "PASP"}

@string{mnras = "Mon. Not. Roy. Astron. Soc."}

@string{aap = "A\&A"}

@string{actaa = "Acta Astron."}

@article{Lu:2020bmd,
    author = "Lu, Philip and Takhistov, Volodymyr and Gelmini, Graciela B. and Hayashi, Kohei and Inoue, Yoshiyuki and Kusenko, Alexander",
    title = "{Constraining Primordial Black Holes with Dwarf Galaxy Heating}",
    eprint = "2007.02213",
    archivePrefix = "arXiv",
    primaryClass = "astro-ph.CO",
    reportNumber = "IPMU20-0076, RIKEN-iTHEMS-Report-20",
    doi = "10.3847/2041-8213/abdcb6",
    journal = "Astrophys. J. Lett.",
    volume = "908",
    number = "2",
    pages = "L23",
    year = "2021"
}

@article{Wadekar:2022ymq,
    author = "Wadekar, Digvijay and Wang, Zihui",
    title = "{Constraining axion and compact dark matter with interstellar medium heating}",
    eprint = "2211.07668",
    archivePrefix = "arXiv",
    primaryClass = "hep-ph",
    doi = "10.1103/PhysRevD.107.083011",
    journal = "Phys. Rev. D",
    volume = "107",
    number = "8",
    pages = "083011",
    year = "2023"
}

@article{Kim:2025gck,
    author = "Kim, TaeHun and Lu, Philip and Takhistov, Volodymyr",
    title = "{Unified Gas Heating Constraints on Extended Dark Matter Compact Objects}",
    eprint = "2508.18344",
    archivePrefix = "arXiv",
    primaryClass = "hep-ph",
    reportNumber = "KEK-QUP-2025-0018, KEK-TH-2748",
    month = "8",
    year = "2025",
    journal=""
}

@article{Takhistov:2021aqx,
    author = "Takhistov, Volodymyr and Lu, Philip and Gelmini, Graciela B. and Hayashi, Kohei and Inoue, Yoshiyuki and Kusenko, Alexander",
    title = "{Interstellar gas heating by primordial black holes}",
    eprint = "2105.06099",
    archivePrefix = "arXiv",
    primaryClass = "astro-ph.GA",
    reportNumber = "IPMU21-0029, RIKEN-iTHEMS-Report-21",
    doi = "10.1088/1475-7516/2022/03/017",
    journal = "JCAP",
    volume = "03",
    number = "03",
    pages = "017",
    year = "2022"
}

@article{Eggemeier:2019khm,
    author = "Eggemeier, Benedikt and Redondo, Javier and Dolag, Klaus and Niemeyer, Jens C. and Vaquero, Alejandro",
    title = "{First Simulations of Axion Minicluster Halos}",
    eprint = "1911.09417",
    archivePrefix = "arXiv",
    primaryClass = "astro-ph.CO",
    doi = "10.1103/PhysRevLett.125.041301",
    journal = "Phys. Rev. Lett.",
    volume = "125",
    number = "4",
    pages = "041301",
    year = "2020"
}

@ARTICLE{2025arXiv250403418B,
       author = {{Brinkerink}, C.~D. and {Arts}, M.~J. and {Bentum}, M.~J. and {Boonstra}, A.~J. and {Cecconi}, B. and {Fialkov}, A. and {Garcia Guti{\'e}rrez}, J. and {Ghosh}, S. and {Grenouilleau}, J. and {Gurvits}, L.~I. and {Klein-Wolt}, M. and {Koopmans}, L.~V.~E. and {Lazendic-Galloway}, J. and {Paragi}, Z. and {Prinsloo}, D. and {Rajan}, R.~T. and {Rouill{\'e}}, E. and {Ruiter}, M. and {Tauber}, J.~A. and {Vedantham}, H.~K. and {Vecchio}, A. and {Vertegaal}, C.~J.~C. and {Zandboer}, J.~C.~F. and {Zucca}, P.},
        title = "{The Dark Ages Explorer (DEX): a filled-aperture ultra-long wavelength radio interferometer on the lunar far side}",
      journal = {arXiv e-prints},
         year = 2025,
        month = apr,
          eid = {arXiv:2504.03418},
        pages = {arXiv:2504.03418},
          doi = {10.48550/arXiv.2504.03418},
archivePrefix = {arXiv},
       eprint = {2504.03418},
 primaryClass = {astro-ph.IM},
       adsurl = {https://ui.adsabs.harvard.edu/abs/2025arXiv250403418B}
}

@article{Burns:2021ndk,
    author = "Burns, Jack and others",
    title = "{Global 21-cm Cosmology from the Farside of the Moon}",
    eprint = "2103.05085",
    journal  = "",
    archivePrefix = "arXiv",
    primaryClass = "astro-ph.CO",
    month = "3",
    year = "2021"
}

@article{Enander:2017ogx,
    author = "Enander, Jonas and Pargner, Andreas and Schwetz, Thomas",
    title = "{Axion minicluster power spectrum and mass function}",
    eprint = "1708.04466",
    archivePrefix = "arXiv",
    primaryClass = "astro-ph.CO",
    doi = "10.1088/1475-7516/2017/12/038",
    journal = "JCAP",
    volume = "12",
    pages = "038",
    year = "2017"
}

@article{Ellis:2020gtq,
    author = "Ellis, David and Marsh, David J. E. and Behrens, Christoph",
    title = "{Axion Miniclusters Made Easy}",
    eprint = "2006.08637",
    archivePrefix = "arXiv",
    primaryClass = "astro-ph.CO",
    doi = "10.1103/PhysRevD.103.083525",
    journal = "Phys. Rev. D",
    volume = "103",
    number = "8",
    pages = "083525",
    year = "2021"
}

@ARTICLE{1990A&A...232..447J,
       author = {{Just}, A. and {Kegel}, W.~H.},
        title = "{Spatial structure and excitation timescales of fluctuations in the interstellar medium induced by the system of stars.}",
      journal = aap,
         year = 1990,
        month = jun,
       volume = {232},
        pages = {447},
       adsurl = {https://ui.adsabs.harvard.edu/abs/1990A&A...232..447J}
}

@article{Murray:2013qza,
    author = "Murray, Steven and Power, Chris and Robotham, A. S. G.",
    title = "{HMFcalc: An online tool for calculating dark matter halo mass functions}",
    eprint = "1306.6721",
    archivePrefix = "arXiv",
    primaryClass = "astro-ph.CO",
    doi = "10.1016/j.ascom.2013.11.001",
    journal = "Astron. Comput.",
    volume = "3-4",
    pages = "23--34",
    year = "2013"
}

@article{Munoz:2019rhi,
    author = "Mu{\~n}oz, Julian B.",
    title = "{Robust Velocity-induced Acoustic Oscillations at Cosmic Dawn}",
    eprint = "1904.07881",
    archivePrefix = "arXiv",
    primaryClass = "astro-ph.CO",
    doi = "10.1103/PhysRevD.100.063538",
    journal = "Phys. Rev. D",
    volume = "100",
    number = "6",
    pages = "063538",
    year = "2019"
}

@ARTICLE{2025MNRAS.tmp.1518H,
       author = {{Hoof}, Sebastian and {Marsh}, David J.~E. and {Sisk-Reyn{\'e}s}, J{\'u}lia and {Matthews}, James H. and {Reynolds}, Christopher},
        title = "{Getting More Out of Black Hole Superradiance: a Statistically Rigorous Approach to Ultralight Boson Constraints from Black Hole Spin Measurements}",
      journal = mnras,
         year = 2025,
        month = sep,
          doi = {10.1093/mnras/staf1564},
archivePrefix = {arXiv},
       eprint = {2406.10337},
 primaryClass = {hep-ph},
       adsurl = {https://ui.adsabs.harvard.edu/abs/2025MNRAS.tmp.1518H}
}

@article{Witte:2024drg,
    author = "Witte, Samuel J. and Mummery, Andrew",
    title = "{Stepping up superradiance constraints on axions}",
    eprint = "2412.03655",
    archivePrefix = "arXiv",
    primaryClass = "hep-ph",
    doi = "10.1103/PhysRevD.111.083044",
    journal = "Phys. Rev. D",
    volume = "111",
    number = "8",
    pages = "083044",
    year = "2025"
}

@article{Agius:2025xbj,
    author = "Agius, Dominic and Essig, Rouven and Gaggero, Daniele and Palomares-Ruiz, Sergio and Suczewski, Gregory and Valli, Mauro",
    title = "{Astrophysical uncertainties challenge 21-cm forecasts: A primordial black hole case study}",
    eprint = "2510.14877",
    archivePrefix = "arXiv",
    primaryClass = "astro-ph.CO",
    month = "10",
    year = "2025",
    journal=""
}

@article{Barkana:2016nyr,
    author = "Barkana, Rennan",
    title = "{The Rise of the First Stars: Supersonic Streaming, Radiative Feedback, and 21-cm Cosmology}",
    eprint = "1605.04357",
    archivePrefix = "arXiv",
    primaryClass = "astro-ph.CO",
    doi = "10.1016/j.physrep.2016.06.006",
    journal = "Phys. Rept.",
    volume = "645",
    pages = "1--59",
    year = "2016"
}

@ARTICLE{2005ApJ...622.1356Z,
       author = {{Zygelman}, B.},
        title = "{Hyperfine Level-changing Collisions of Hydrogen Atoms and Tomography of the Dark Age Universe}",
      journal = apj,
         year = 2005,
        month = apr,
       volume = {622},
       number = {2},
        pages = {1356-1362},
          doi = {10.1086/427682},
       adsurl = {https://ui.adsabs.harvard.edu/abs/2005ApJ...622.1356Z}
}

@article{Furlanetto:2006su,
    author = "Furlanetto, Steven and Furlanetto, Michael",
    title = "{Spin Exchange Rates in Electron-Hydrogen Collisions}",
    eprint = "astro-ph/0608067",
    archivePrefix = "arXiv",
    doi = "10.1111/j.1365-2966.2006.11169.x",
    journal = "Mon. Not. Roy. Astron. Soc.",
    volume = "374",
    pages = "547--555",
    year = "2007"
}

@article{Furlanetto:2007te,
    author = "Furlanetto, Steven and Furlanetto, Michael",
    title = "{Spin Exchange Rates in Proton-Hydrogen Collisions}",
    eprint = "astro-ph/0702487",
    archivePrefix = "arXiv",
    doi = "10.1111/j.1365-2966.2007.11921.x",
    journal = "Mon. Not. Roy. Astron. Soc.",
    volume = "379",
    pages = "130--134",
    year = "2007"
}

@article{Munoz:2023kkg,
    author = "Mu{\~n}oz, Julian B.",
    title = "{An effective model for the cosmic-dawn 21-cm signal}",
    eprint = "2302.08506",
    archivePrefix = "arXiv",
    primaryClass = "astro-ph.CO",
    doi = "10.1093/mnras/stad1512",
    journal = "Mon. Not. Roy. Astron. Soc.",
    volume = "523",
    number = "2",
    pages = "2587--2607",
    year = "2023"
}

@article{Reis:2021nqf,
    author = "Reis, Itamar and Fialkov, Anastasia and Barkana, Rennan",
    title = "{The subtlety of Ly{\,}{\ensuremath{\alpha}} photons: changing the expected range of the 21-cm signal}",
    eprint = "2101.01777",
    archivePrefix = "arXiv",
    primaryClass = "astro-ph.CO",
    doi = "10.1093/mnras/stab2089",
    journal = "Mon. Not. Roy. Astron. Soc.",
    volume = "506",
    number = "4",
    pages = "5479--5493",
    year = "2021"
}

@article{Liu:2019bbm,
    author = "Liu, Hongwan and Ridgway, Gregory W. and Slatyer, Tracy R.",
    title = "{Code package for calculating modified cosmic ionization and thermal histories with dark matter and other exotic energy injections}",
    eprint = "1904.09296",
    archivePrefix = "arXiv",
    primaryClass = "astro-ph.CO",
    doi = "10.1103/PhysRevD.101.023530",
    journal = "Phys. Rev. D",
    volume = "101",
    number = "2",
    pages = "023530",
    year = "2020"
}

@article{Facchinetti:2023slb,
    author = "Facchinetti, Ga{\'e}tan and Lopez-Honorez, Laura and Qin, Yuxiang and Mesinger, Andrei",
    title = "{21cm signal sensitivity to dark matter decay}",
    eprint = "2308.16656",
    archivePrefix = "arXiv",
    primaryClass = "astro-ph.CO",
    reportNumber = "ULB-TH/23-09",
    doi = "10.1088/1475-7516/2024/01/005",
    journal = "JCAP",
    volume = "01",
    pages = "005",
    year = "2024"
}

@article{Sun:2023acy,
    author = "Sun, Yitian and Foster, Joshua W. and Liu, Hongwan and Mu{\~n}oz, Julian B. and Slatyer, Tracy R.",
    title = "{Inhomogeneous energy injection in the 21-cm power spectrum: Sensitivity to dark matter decay}",
    eprint = "2312.11608",
    archivePrefix = "arXiv",
    primaryClass = "hep-ph",
    reportNumber = "MIT-CTP/5657, FERMILAB-PUB-23-0816-T-V",
    doi = "10.1103/PhysRevD.111.043015",
    journal = "Phys. Rev. D",
    volume = "111",
    number = "4",
    pages = "043015",
    year = "2025"
}

@article{Sun:2025ksr,
    author = "Sun, Yitian and Foster, Joshua W. and Mu{\~n}oz, Julian B.",
    title = "{Constraining inhomogeneous energy injection from annihilating dark matter and primordial black holes with 21-cm cosmology}",
    eprint = "2509.22772",
    archivePrefix = "arXiv",
    primaryClass = "hep-ph",
    month = "9",
    year = "2025",
    journal=""
}

@article{Chen:2003gc,
    author = "Chen, Xue-Lei and Miralda-Escude, Jordi",
    title = "{The spin - kinetic temperature coupling and the heating rate due to Lyman - alpha scattering before reionization: Predictions for 21cm emission and absorption}",
    eprint = "astro-ph/0303395",
    archivePrefix = "arXiv",
    doi = "10.1086/380829",
    journal = "Astrophys. J.",
    volume = "602",
    pages = "1--11",
    year = "2004"
}

@article{Chuzhoy:2005wv,
    author = "Chuzhoy, Leonid and Shapiro, Paul R.",
    title = "{UV pumping of hyperfine transitions in the light elements, with application to 21-cm hydrogen and 92-cm deuterium lines from the early universe}",
    eprint = "astro-ph/0512206",
    archivePrefix = "arXiv",
    doi = "10.1086/507670",
    journal = "Astrophys. J.",
    volume = "651",
    pages = "1--7",
    year = "2006"
}

@article{Venumadhav:2018uwn,
    author = "Venumadhav, Tejaswi and Dai, Liang and Kaurov, Alexander and Zaldarriaga, Matias",
    title = "{Heating of the intergalactic medium by the cosmic microwave background during cosmic dawn}",
    eprint = "1804.02406",
    archivePrefix = "arXiv",
    primaryClass = "astro-ph.CO",
    doi = "10.1103/PhysRevD.98.103513",
    journal = "Phys. Rev. D",
    volume = "98",
    number = "10",
    pages = "103513",
    year = "2018"
}

@article{Furlanetto:2006fs,
    author = "Furlanetto, Steven and Pritchard, Jonathan R.",
    title = "{The Scattering of Lyman-series Photons in the Intergalactic Medium}",
    eprint = "astro-ph/0605680",
    archivePrefix = "arXiv",
    doi = "10.1111/j.1365-2966.2006.10899.x",
    journal = "Mon. Not. Roy. Astron. Soc.",
    volume = "372",
    pages = "1093--1103",
    year = "2006"
}

@article{Xu:2018efh,
    author = "Xu, Weishuang Linda and Dvorkin, Cora and Chael, Andrew",
    title = "{Probing sub-GeV Dark Matter-Baryon Scattering with Cosmological Observables}",
    eprint = "1802.06788",
    archivePrefix = "arXiv",
    primaryClass = "astro-ph.CO",
    doi = "10.1103/PhysRevD.97.103530",
    journal = "Phys. Rev. D",
    volume = "97",
    number = "10",
    pages = "103530",
    year = "2018"
}

@article{Munoz:2018jwq,
    author = "Mu{\~n}oz, Julian B. and Dvorkin, Cora and Loeb, Abraham",
    title = "{21-cm Fluctuations from Charged Dark Matter}",
    eprint = "1804.01092",
    archivePrefix = "arXiv",
    primaryClass = "astro-ph.CO",
    doi = "10.1103/PhysRevLett.121.121301",
    journal = "Phys. Rev. Lett.",
    volume = "121",
    number = "12",
    pages = "121301",
    year = "2018"
}

@article{Munoz:2015bca,
    author = {Mu{\~n}oz, Julian B. and Kovetz, Ely D. and Ali-Ha{\"\i}moud, Yacine},
    title = "{Heating of Baryons due to Scattering with Dark Matter During the Dark Ages}",
    eprint = "1509.00029",
    archivePrefix = "arXiv",
    primaryClass = "astro-ph.CO",
    doi = "10.1103/PhysRevD.92.083528",
    journal = "Phys. Rev. D",
    volume = "92",
    number = "8",
    pages = "083528",
    year = "2015"
}

@article{Munoz:2017qpy,
    author = "Mu{\~n}oz, Julian B. and Loeb, Abraham",
    title = "{Constraints on Dark Matter-Baryon Scattering from the Temperature Evolution of the Intergalactic Medium}",
    eprint = "1708.08923",
    archivePrefix = "arXiv",
    primaryClass = "astro-ph.CO",
    doi = "10.1088/1475-7516/2017/11/043",
    journal = "JCAP",
    volume = "11",
    pages = "043",
    year = "2017"
}

@article{Barkana:2018lgd,
    author = "Barkana, Rennan",
    title = "{Possible interaction between baryons and dark-matter particles revealed by the first stars}",
    eprint = "1803.06698",
    archivePrefix = "arXiv",
    primaryClass = "astro-ph.CO",
    doi = "10.1038/nature25791",
    journal = "Nature",
    volume = "555",
    number = "7694",
    pages = "71--74",
    year = "2018"
}

@article{Barkana:2018qrx,
    author = "Barkana, Rennan and Outmezguine, Nadav Joseph and Redigolo, Diego and Volansky, Tomer",
    title = "{Strong constraints on light dark matter interpretation of the EDGES signal}",
    eprint = "1803.03091",
    archivePrefix = "arXiv",
    primaryClass = "hep-ph",
    doi = "10.1103/PhysRevD.98.103005",
    journal = "Phys. Rev. D",
    volume = "98",
    number = "10",
    pages = "103005",
    year = "2018"
}

@article{Fialkov:2018xre,
    author = "Fialkov, Anastasia and Barkana, Rennan and Cohen, Aviad",
    title = "{Constraining Baryon--Dark Matter Scattering with the Cosmic Dawn 21-cm Signal}",
    eprint = "1802.10577",
    archivePrefix = "arXiv",
    primaryClass = "astro-ph.CO",
    doi = "10.1103/PhysRevLett.121.011101",
    journal = "Phys. Rev. Lett.",
    volume = "121",
    pages = "011101",
    year = "2018"
}

@article{Munoz:2018pzp,
    author = "Mu{\~n}oz, Julian B. and Loeb, Abraham",
    title = "{A small amount of mini-charged dark matter could cool the baryons in the early Universe}",
    eprint = "1802.10094",
    archivePrefix = "arXiv",
    primaryClass = "astro-ph.CO",
    doi = "10.1038/s41586-018-0151-x",
    journal = "Nature",
    volume = "557",
    number = "7707",
    pages = "684",
    year = "2018"
}

@article{DES:2020fxi,
    author = "Nadler, E. O. and others",
    collaboration = "DES",
    title = "{Milky Way Satellite Census. III. Constraints on Dark Matter Properties from Observations of Milky Way Satellite Galaxies}",
    eprint = "2008.00022",
    archivePrefix = "arXiv",
    primaryClass = "astro-ph.CO",
    reportNumber = "FERMILAB-PUB-20-277-AE, SLAC-PUB-17554, DES-2020-546",
    doi = "10.1103/PhysRevLett.126.091101",
    journal = "Phys. Rev. Lett.",
    volume = "126",
    pages = "091101",
    year = "2021"
}

@article{Nguyen:2021cnb,
    author = "Nguyen, David V. and Sarnaaik, Dimple and Boddy, Kimberly K. and Nadler, Ethan O. and Gluscevic, Vera",
    title = "{Observational constraints on dark matter scattering with electrons}",
    eprint = "2107.12380",
    archivePrefix = "arXiv",
    primaryClass = "astro-ph.CO",
    reportNumber = "UTTG-04-2021",
    doi = "10.1103/PhysRevD.104.103521",
    journal = "Phys. Rev. D",
    volume = "104",
    number = "10",
    pages = "103521",
    year = "2021"
}

@article{Dalal:2010yt,
    author = "Dalal, Neal and Pen, Ue-Li and Seljak, Uros",
    title = "{Large-scale BAO signatures of the smallest galaxies}",
    eprint = "1009.4704",
    archivePrefix = "arXiv",
    primaryClass = "astro-ph.CO",
    doi = "10.1088/1475-7516/2010/11/007",
    journal = "JCAP",
    volume = "11",
    pages = "007",
    year = "2010"
}

@article{Barkana:2022hko,
    author = "Barkana, Rennan and Fialkov, Anastasia and Liu, Hongwan and Outmezguine, Nadav Joseph",
    title = "{Anticipating a new physics signal in upcoming 21-cm power spectrum observations}",
    eprint = "2212.08082",
    archivePrefix = "arXiv",
    primaryClass = "hep-ph",
    doi = "10.1103/PhysRevD.108.063503",
    journal = "Phys. Rev. D",
    volume = "108",
    number = "6",
    pages = "063503",
    year = "2023"
}

@article{Ali-Haimoud:2013hpa,
    author = {Ali-Ha{\"\i}moud, Yacine and Meerburg, P. Daniel and Yuan, Sihan},
    title = "{New light on 21 cm intensity fluctuations from the dark ages}",
    eprint = "1312.4948",
    archivePrefix = "arXiv",
    primaryClass = "astro-ph.CO",
    doi = "10.1103/PhysRevD.89.083506",
    journal = "Phys. Rev. D",
    volume = "89",
    number = "8",
    pages = "083506",
    year = "2014"
}

@article{Eroncel:2022vjg,
    author = {Er{\"o}ncel, Cem and Sato, Ryosuke and Servant, Geraldine and S{\o}rensen, Philip},
    title = "{ALP dark matter from kinetic fragmentation: opening up the parameter window}",
    eprint = "2206.14259",
    archivePrefix = "arXiv",
    primaryClass = "hep-ph",
    reportNumber = "DESY 22-106, OU-HET-1148",
    doi = "10.1088/1475-7516/2022/10/053",
    journal = "JCAP",
    volume = "10",
    pages = "053",
    year = "2022"
}

@article{Arvanitaki:2019rax,
    author = "Arvanitaki, Asimina and Dimopoulos, Savas and Galanis, Marios and Lehner, Luis and Thompson, Jedidiah O. and Van Tilburg, Ken",
    title = "{Large-misalignment mechanism for the formation of compact axion structures: Signatures from the QCD axion to fuzzy dark matter}",
    eprint = "1909.11665",
    archivePrefix = "arXiv",
    primaryClass = "astro-ph.CO",
    doi = "10.1103/PhysRevD.101.083014",
    journal = "Phys. Rev. D",
    volume = "101",
    number = "8",
    pages = "083014",
    year = "2020"
}

@article{Fox:2025tqa,
    author = "Fox, Patrick J. and Weiner, Neal and Xiao, Huangyu",
    title = "{Radio Killed the Axion Star: Constraining Axion Properties with Radio Telescopes}",
    eprint = "2508.08371",
    archivePrefix = "arXiv",
    primaryClass = "hep-ph",
    reportNumber = "FERMILAB-PUB-25-0567-T",
    journal = "",
    month = "8",
    year = "2025"
}

@article{Ostriker_1999,
   title={Dynamical Friction in a Gaseous Medium},
   volume={513},
   ISSN={1538-4357},
   url={http://dx.doi.org/10.1086/306858},
   DOI={10.1086/306858},
   number={1},
   journal={The Astrophysical Journal},
   publisher={American Astronomical Society},
   author={Ostriker, Eve C.},
   year={1999},
   month=mar, pages={252–258} }

@article{Graham:2024hah,
    author = "Graham, Peter W. and Ramani, Harikrishnan",
    title = "{Constraints on dark matter from dynamical heating of stars in ultrafaint dwarfs. II. Substructure and the primordial power spectrum}",
    eprint = "2404.01378",
    archivePrefix = "arXiv",
    primaryClass = "hep-ph",
    doi = "10.1103/PhysRevD.110.075012",
    journal = "Phys. Rev. D",
    volume = "110",
    number = "7",
    pages = "075012",
    year = "2024"
}

@article{Chang:2024fol,
    author = "Chang, Jae Hyeok and Fox, Patrick J. and Xiao, Huangyu",
    title = "{Axion stars: mass functions and constraints}",
    eprint = "2406.09499",
    archivePrefix = "arXiv",
    primaryClass = "hep-ph",
    reportNumber = "FERMILAB-PUB-24-0295-T",
    doi = "10.1088/1475-7516/2024/08/023",
    journal = "JCAP",
    volume = "08",
    pages = "023",
    year = "2024"
}

@article{Ali-Haimoud:2016,
    author = {Ali-Ha{\"i}moud, Yacine and Kamionkowski, Marc},
    title = "{Cosmic microwave background limits on accreting primordial black holes}",
    eprint = "1612.05644",
    archivePrefix = "arXiv",
    primaryClass = "astro-ph.CO",
    doi = "10.1103/PhysRevD.95.043534",
    journal = "Phys. Rev. D",
    volume = "95",
    number = "4",
    pages = "043534",
    year = "2017"
}

@article{Brandt:2016aco,
    author = "Brandt, Timothy D.",
    title = "{Constraints on MACHO Dark Matter from Compact Stellar Systems in Ultra-Faint Dwarf Galaxies}",
    eprint = "1605.03665",
    archivePrefix = "arXiv",
    primaryClass = "astro-ph.GA",
    doi = "10.3847/2041-8205/824/2/L31",
    journal = "Astrophys. J. Lett.",
    volume = "824",
    number = "2",
    pages = "L31",
    year = "2016"
}

@article{Graham:2023unf,
    author = "Graham, Peter W. and Ramani, Harikrishnan",
    title = "{Constraints on Dark Matter from Dynamical Heating of Stars in Ultrafaint Dwarfs. Part 1: MACHOs and Primordial Black Holes}",
    eprint = "2311.07654",
    archivePrefix = "arXiv",
    primaryClass = "hep-ph",
    journal = "",
    month = "11",
    year = "2023"
}

@article{Fox:2023xgx,
    author = "Fox, Patrick J. and Weiner, Neal and Xiao, Huangyu",
    title = "{Recurrent axion stars collapse with dark radiation emission and their cosmological constraints}",
    eprint = "2302.00685",
    archivePrefix = "arXiv",
    primaryClass = "hep-ph",
    reportNumber = "FERMILAB-PUB-23-029-T",
    doi = "10.1103/PhysRevD.108.095043",
    journal = "Phys. Rev. D",
    volume = "108",
    number = "9",
    pages = "095043",
    year = "2023"
}

@article{Escudero:2023vgv,
    author = "Escudero, Miguel and Pooni, Charis Kaur and Fairbairn, Malcolm and Blas, Diego and Du, Xiaolong and Marsh, David J. E.",
    title = "{Axion Star Explosions: A New Source for Axion Indirect Detection}",
    eprint = "2302.10206",
    archivePrefix = "arXiv",
    primaryClass = "hep-ph",
    reportNumber = "KCL-PH-TH-2023-16, CERN-TH-2023-029",
    journal = "",
    month = "2",
    year = "2023"
}

@ARTICLE{1974ApJ...187..425P,
	author = {{Press}, William H. and {Schechter}, Paul},
	title = "{Formation of Galaxies and Clusters of Galaxies by Self-Similar Gravitational Condensation}",
	journal = apj,
	year = 1974,
	month = feb,
	volume = {187},
	pages = {425-438},
	doi = {10.1086/152650},
	adsurl = {https://ui.adsabs.harvard.edu/abs/1974ApJ...187..425P}
}

@article{Mehta:2020kwu,
    author = "Mehta, Viraf M. and Demirtas, Mehmet and Long, Cody and Marsh, David J. E. and Mcallister, Liam and Stott, Matthew J.",
    title = "{Superradiance Exclusions in the Landscape of Type IIB String Theory}",
    eprint = "2011.08693",
    archivePrefix = "arXiv",
    primaryClass = "hep-th",
    reportNumber = "KCL-PH-TH/2020-77",
    month = "11",
    year = "2020",
    journal = ""
}

@article{Baryakhtar:2020gao,
    author = "Baryakhtar, Masha and Galanis, Marios and Lasenby, Robert and Simon, Olivier",
    title = "{Black hole superradiance of self-interacting scalar fields}",
    eprint = "2011.11646",
    archivePrefix = "arXiv",
    primaryClass = "hep-ph",
    doi = "10.1103/PhysRevD.103.095019",
    journal = "Phys. Rev. D",
    volume = "103",
    number = "9",
    pages = "095019",
    year = "2021"
}

@article{Hogan:1988mp,
    author = "Hogan, C. J. and Rees, M. J.",
    title = "{Axion Miniclusters}",
    doi = "10.1016/0370-2693(88)91655-3",
    journal = "Phys. Lett. B",
    volume = "205",
    pages = "228--230",
    year = "1988"
}

@article{Kolb:1993zz,
    author = "Kolb, Edward W. and Tkachev, Igor I.",
    title = "{Axion miniclusters and Bose stars}",
    eprint = "hep-ph/9303313",
    archivePrefix = "arXiv",
    reportNumber = "FERMILAB-PUB-93-066-A",
    doi = "10.1103/PhysRevLett.71.3051",
    journal = "Phys. Rev. Lett.",
    volume = "71",
    pages = "3051--3054",
    year = "1993"
}

@article{Kolb:1995bu,
    author = "Kolb, Edward W. and Tkachev, Igor I.",
    title = "{Femtolensing and picolensing by axion miniclusters}",
    eprint = "astro-ph/9510043",
    archivePrefix = "arXiv",
    reportNumber = "FERMILAB-PUB-95-309-A, OSU-TA-20-95",
    doi = "10.1086/309962",
    journal = "Astrophys. J. Lett.",
    volume = "460",
    pages = "L25--L28",
    year = "1996"
}

@article{Eroncel:2022efc,
    author = {Er\"oncel, Cem and Servant, G\'eraldine},
    title = "{ALP Dark Matter Mini-Clusters from Kinetic Fragmentation}",
    eprint = "2207.10111",
    archivePrefix = "arXiv",
    primaryClass = "hep-ph",
    reportNumber = "DESY 22-115",
    journal = "",
    month = "7",
    year = "2022"
}

@article{Co:2019jts,
    author = "Co, Raymond T. and Hall, Lawrence J. and Harigaya, Keisuke",
    title = "{Axion Kinetic Misalignment Mechanism}",
    eprint = "1910.14152",
    archivePrefix = "arXiv",
    primaryClass = "hep-ph",
    reportNumber = "LCTP-19-28",
    doi = "10.1103/PhysRevLett.124.251802",
    journal = "Phys. Rev. Lett.",
    volume = "124",
    number = "25",
    pages = "251802",
    year = "2020"
}

@article{Xiao:2021nkb,
    author = "Xiao, Huangyu and Williams, Ian and McQuinn, Matthew",
    title = "{Simulations of axion minihalos}",
    eprint = "2101.04177",
    archivePrefix = "arXiv",
    primaryClass = "astro-ph.CO",
    doi = "10.1103/PhysRevD.104.023515",
    journal = "Phys. Rev. D",
    volume = "104",
    number = "2",
    pages = "023515",
    year = "2021"
}

@article{Ruffini:1969qy,
    author = "Ruffini, Remo and Bonazzola, Silvano",
    title = "{Systems of selfgravitating particles in general relativity and the concept of an equation of state}",
    doi = "10.1103/PhysRev.187.1767",
    journal = "Phys. Rev.",
    volume = "187",
    pages = "1767--1783",
    year = "1969"
}

@misc{AxionLimits,
  author       = {Ciaran O'Hare},
  title        = {cajohare/AxionLimits: AxionLimits},
  month        = jul,
  year         = 2020,
  publisher    = {Zenodo},
  version      = {v1.0},
  doi          = {10.5281/zenodo.3932430},
  howpublished = {\url{https://cajohare.github.io/AxionLimits/}}
}

@article{Unal:2020jiy,
    author = {\"Unal, Caner and Pacucci, Fabio and Loeb, Abraham},
    title = "{Properties of ultralight bosons from heavy quasar spins via superradiance}",
    eprint = "2012.12790",
    archivePrefix = "arXiv",
    primaryClass = "hep-ph",
    doi = "10.1088/1475-7516/2021/05/007",
    journal = "JCAP",
    volume = "05",
    pages = "007",
    year = "2021"
}

@article{Graham:2025opw,
    author = "Graham, Peter W. and Ramani, Harikrishnan and Ruhdorfer, Maximilian",
    title = "{Robust bounds on MACHOs from the faintest galaxies}",
    eprint = "2510.01310",
    archivePrefix = "arXiv",
    primaryClass = "hep-ph",
    journal = "",
    month = "10",
    year = "2025"
}

@article{EROS-2:2006ryy,
    author = "Tisserand, P. and others",
    collaboration = "EROS-2",
    title = "{Limits on the Macho Content of the Galactic Halo from the EROS-2 Survey of the Magellanic Clouds}",
    eprint = "astro-ph/0607207",
    archivePrefix = "arXiv",
    doi = "10.1051/0004-6361:20066017",
    journal = "Astron. Astrophys.",
    volume = "469",
    pages = "387--404",
    year = "2007"
}

@article{Blaineau:2022nhy,
    author = "Blaineau, T. and others",
    title = "{New limits from microlensing on Galactic black holes in the mass range 10 M{\ensuremath{\odot}} {\ensuremath{<}} M {\ensuremath{<}} 1000 M{\ensuremath{\odot}}}",
    eprint = "2202.13819",
    archivePrefix = "arXiv",
    primaryClass = "astro-ph.GA",
    doi = "10.1051/0004-6361/202243430",
    journal = "Astron. Astrophys.",
    volume = "664",
    pages = "A106",
    year = "2022"
}

@article{MACHO:2000qbb,
    author = "Alcock, C. and others",
    collaboration = "MACHO",
    title = "{The MACHO project: Microlensing results from 5.7 years of LMC observations}",
    eprint = "astro-ph/0001272",
    archivePrefix = "arXiv",
    doi = "10.1086/309512",
    journal = "Astrophys. J.",
    volume = "542",
    pages = "281--307",
    year = "2000"
}

@ARTICLE{OGLE:2015,
    author = {{Udalski}, A. and {Szyma{\'n}ski}, M.~K. and {Szyma{\'n}ski}, G.},
    title = "{OGLE-IV: Fourth Phase of the Optical Gravitational Lensing Experiment}",
    journal = actaa,
    year = 2015,
    month = mar,
    volume = {65},
    number = {1},
    pages = {1-38},
    doi = {10.48550/arXiv.1504.05966},
    archivePrefix = {arXiv},
    eprint = {1504.05966},
    primaryClass = {astro-ph.SR},
    adsurl = {https://ui.adsabs.harvard.edu/abs/2015AcA....65....1U}
}

@article{Green:2020jor,
    author = "Green, Anne M. and Kavanagh, Bradley J.",
    title = "{Primordial Black Holes as a dark matter candidate}",
    eprint = "2007.10722",
    archivePrefix = "arXiv",
    primaryClass = "astro-ph.CO",
    doi = "10.1088/1361-6471/abc534",
    journal = "J. Phys. G",
    volume = "48",
    number = "4",
    pages = "043001",
    year = "2021"
}

@article{Carr:2020xqk,
    author = "Carr, Bernard and Kuhnel, Florian",
    title = "{Primordial Black Holes as Dark Matter: Recent Developments}",
    eprint = "2006.02838",
    archivePrefix = "arXiv",
    primaryClass = "astro-ph.CO",
    doi = "10.1146/annurev-nucl-050520-125911",
    journal = "Ann. Rev. Nucl. Part. Sci.",
    volume = "70",
    pages = "355--394",
    year = "2020"
}

@article{Ricotti:2007au,
    author = "Ricotti, Massimo and Ostriker, Jeremiah P. and Mack, Katherine J.",
    title = "{Effect of Primordial Black Holes on the Cosmic Microwave Background and Cosmological Parameter Estimates}",
    eprint = "0709.0524",
    archivePrefix = "arXiv",
    primaryClass = "astro-ph",
    doi = "10.1086/587831",
    journal = "Astrophys. J.",
    volume = "680",
    pages = "829",
    year = "2008"
}

@article{Bai:2018dxf,
    author = "Bai, Yang and Long, Andrew J. and Lu, Sida",
    title = "{Dark Quark Nuggets}",
    eprint = "1810.04360",
    archivePrefix = "arXiv",
    primaryClass = "hep-ph",
    reportNumber = "FERMILAB-PUB-18-600-T",
    doi = "10.1103/PhysRevD.99.055047",
    journal = "Phys. Rev. D",
    volume = "99",
    number = "5",
    pages = "055047",
    year = "2019"
}

@article{Witten:1984rs,
    author = "Witten, Edward",
    title = "{Cosmic Separation of Phases}",
    reportNumber = "PRINT-84-0400 (IAS,PRINCETON)",
    doi = "10.1103/PhysRevD.30.272",
    journal = "Phys. Rev. D",
    volume = "30",
    pages = "272--285",
    year = "1984"
}

@article{Kusenko:1997si,
    author = "Kusenko, Alexander and Shaposhnikov, Mikhail E.",
    title = "{Supersymmetric Q balls as dark matter}",
    eprint = "hep-ph/9709492",
    archivePrefix = "arXiv",
    reportNumber = "CERN-TH-97-259",
    doi = "10.1016/S0370-2693(97)01375-0",
    journal = "Phys. Lett. B",
    volume = "418",
    pages = "46--54",
    year = "1998"
}

@article{Hong:2020est,
    author = "Hong, Jeong-Pyong and Jung, Sunghoon and Xie, Ke-Pan",
    title = "{Fermi-ball dark matter from a first-order phase transition}",
    eprint = "2008.04430",
    archivePrefix = "arXiv",
    primaryClass = "hep-ph",
    doi = "10.1103/PhysRevD.102.075028",
    journal = "Phys. Rev. D",
    volume = "102",
    number = "7",
    pages = "075028",
    year = "2020"
}

@article{Liebling:2012fv,
    author = "Liebling, Steven L. and Palenzuela, Carlos",
    title = "{Dynamical boson stars}",
    eprint = "1202.5809",
    archivePrefix = "arXiv",
    primaryClass = "gr-qc",
    doi = "10.1007/s41114-023-00043-4",
    journal = "Living Rev. Rel.",
    volume = "26",
    number = "1",
    pages = "1",
    year = "2023"
}

@article{Tseliakhovich:2010bj,
    author = "Tseliakhovich, Dmitriy and Hirata, Christopher",
    title = "{Relative velocity of dark matter and baryonic fluids and the formation of the first structures}",
    eprint = "1005.2416",
    archivePrefix = "arXiv",
    primaryClass = "astro-ph.CO",
    doi = "10.1103/PhysRevD.82.083520",
    journal = "Phys. Rev. D",
    volume = "82",
    pages = "083520",
    year = "2010"
}

@article{Dvorkin:2013cea,
    author = "Dvorkin, Cora and Blum, Kfir and Kamionkowski, Marc",
    title = "{Constraining Dark Matter-Baryon Scattering with Linear Cosmology}",
    eprint = "1311.2937",
    archivePrefix = "arXiv",
    primaryClass = "astro-ph.CO",
    doi = "10.1103/PhysRevD.89.023519",
    journal = "Phys. Rev. D",
    volume = "89",
    number = "2",
    pages = "023519",
    year = "2014"
}

@article{Boddy:2018wzy,
    author = "Boddy, Kimberly K. and Gluscevic, Vera and Poulin, Vivian and Kovetz, Ely D. and Kamionkowski, Marc and Barkana, Rennan",
    title = "{Critical assessment of CMB limits on dark matter-baryon scattering: New treatment of the relative bulk velocity}",
    eprint = "1808.00001",
    archivePrefix = "arXiv",
    primaryClass = "astro-ph.CO",
    doi = "10.1103/PhysRevD.98.123506",
    journal = "Phys. Rev. D",
    volume = "98",
    number = "12",
    pages = "123506",
    year = "2018"
}

@article{Slatyer:2018aqg,
    author = "Slatyer, Tracy R. and Wu, Chih-Liang",
    title = "{Early-Universe constraints on dark matter-baryon scattering and their implications for a global 21 cm signal}",
    eprint = "1803.09734",
    archivePrefix = "arXiv",
    primaryClass = "astro-ph.CO",
    reportNumber = "MIT-CTP/4995, MIT-CTP-4995",
    doi = "10.1103/PhysRevD.98.023013",
    journal = "Phys. Rev. D",
    volume = "98",
    number = "2",
    pages = "023013",
    year = "2018"
}

@article{Buen-Abad:2021mvc,
    author = "Buen-Abad, Manuel A. and Essig, Rouven and McKeen, David and Zhong, Yi-Ming",
    title = "{Cosmological constraints on dark matter interactions with ordinary matter}",
    eprint = "2107.12377",
    archivePrefix = "arXiv",
    primaryClass = "astro-ph.CO",
    doi = "10.1016/j.physrep.2022.02.006",
    journal = "Phys. Rept.",
    volume = "961",
    pages = "1--35",
    year = "2022"
}

@article{Evoli:2014pva,
    author = "Evoli, Carmelo and Mesinger, Andrei and Ferrara, Andrea",
    title = "{Unveiling the nature of dark matter with high redshift 21 cm line experiments}",
    eprint = "1408.1109",
    archivePrefix = "arXiv",
    primaryClass = "astro-ph.HE",
    doi = "10.1088/1475-7516/2014/11/024",
    journal = "JCAP",
    volume = "11",
    pages = "024",
    year = "2014"
}

@article{Lopez-Honorez:2016sur,
    author = "Lopez-Honorez, Laura and Mena, Olga and Molin{\'e}, {\'A}ngeles and Palomares-Ruiz, Sergio and Vincent, Aaron C.",
    title = "{The 21 cm signal and the interplay between dark matter annihilations and astrophysical processes}",
    eprint = "1603.06795",
    archivePrefix = "arXiv",
    primaryClass = "astro-ph.CO",
    reportNumber = "CFTP-16-007, IFIC-16-16, IPPP-16-20",
    doi = "10.1088/1475-7516/2016/08/004",
    journal = "JCAP",
    volume = "08",
    pages = "004",
    year = "2016"
}

@article{Liu:2019knx,
    author = "Liu, Hongwan and Outmezguine, Nadav Joseph and Redigolo, Diego and Volansky, Tomer",
    title = "{Reviving Millicharged Dark Matter for 21-cm Cosmology}",
    eprint = "1908.06986",
    archivePrefix = "arXiv",
    primaryClass = "hep-ph",
    reportNumber = "MIT-CTP/5126",
    doi = "10.1103/PhysRevD.100.123011",
    journal = "Phys. Rev. D",
    volume = "100",
    number = "12",
    pages = "123011",
    year = "2019"
}

@ARTICLE{Philip:2019,
    author = {{Philip}, L. and {Abdurashidova}, Z. and {Chiang}, H.~C. and {Ghazi}, N. and {Gumba}, A. and {Heilgendorff}, H.~M. and {J{\'a}uregui-Garc{\'\i}a}, J.~M. and {Malepe}, K. and {Nunhokee}, C.~D. and {Peterson}, J. and {Sievers}, J.~L. and {Simes}, V. and {Spann}, R.},
    title = "{Probing Radio Intensity at High-Z from Marion: 2017 Instrument}",
    journal = {Journal of Astronomical Instrumentation},
    year = 2019,
    month = jan,
    volume = {8},
    number = {2},
    eid = {1950004},
    pages = {1950004},
    doi = {10.1142/S2251171719500041},
    archivePrefix = {arXiv},
    eprint = {1806.09531},
    primaryClass = {astro-ph.IM},
    adsurl = {https://ui.adsabs.harvard.edu/abs/2019JAI.....850004P}
}

@article{Voytek:2013nua,
    author = "Voytek, Tabitha C. and Natarajan, Aravind and J{\'a}uregui Garc{\'\i}a, Jos{\'e} Miguel and Peterson, Jeffrey B. and L{\'o}pez-Cruz, Omar",
    title = "{Probing the Dark Ages at $z \sim$ 20: The SCI-HI 21 cm All-sky Spectrum Experiment}",
    eprint = "1311.0014",
    archivePrefix = "arXiv",
    primaryClass = "astro-ph.CO",
    doi = "10.1088/2041-8205/782/1/L9",
    journal = "Astrophys. J. Lett.",
    volume = "782",
    pages = "L9",
    year = "2014"
}

@article{deLeraAcedo:2022kiu,
    author = "de Lera Acedo, E. and others",
    title = "{The REACH radiometer for detecting the 21-cm hydrogen signal from redshift z{\,}{\ensuremath{\approx}}{\,}7.5{\textendash}28}",
    eprint = "2210.07409",
    archivePrefix = "arXiv",
    primaryClass = "astro-ph.CO",
    doi = "10.1038/s41550-022-01817-6",
    journal = "Nature Astron.",
    volume = "6",
    number = "7",
    pages = "998",
    year = "2022"
}

@ARTICLE{HeraII,
       author = "Berkhout, Lindsay M. and others",
        title = "{Hydrogen Epoch of Reionization Array (HERA) Phase II Deployment and Commissioning}",
      journal = pasp,
         year = 2024,
        month = apr,
       volume = {136},
       number = {4},
          eid = {045002},
        pages = {045002},
          doi = {10.1088/1538-3873/ad3122},
archivePrefix = {arXiv},
       eprint = {2401.04304},
 primaryClass = {astro-ph.IM},
       adsurl = {https://ui.adsabs.harvard.edu/abs/2024PASP..136d5002B}
}

@article{SKA,
    author = "Weltman, A. and others",
    title = "{Fundamental physics with the Square Kilometre Array}",
    eprint = "1810.02680",
    archivePrefix = "arXiv",
    primaryClass = "astro-ph.CO",
    doi = "10.1017/pasa.2019.42",
    journal = "Publ. Astron. Soc. Austral.",
    volume = "37",
    pages = "e002",
    year = "2020"
}

@ARTICLE{SKAlow,
       author = {{Labate}, Maria G. and {Waterson}, Mark and {Alachkar}, Bassem and {Hendre}, Aniket and {Lewis}, Peter and {Bartolini}, Marco and {Dewdney}, Peter},
        title = "{Highlights of the Square Kilometre Array Low Frequency (SKA-LOW) Telescope}",
      journal = {Journal of Astronomical Telescopes, Instruments, and Systems},
         year = 2022,
        month = jan,
       volume = {8},
          eid = {011024},
        pages = {011024},
          doi = {10.1117/1.JATIS.8.1.011024},
       adsurl = {https://ui.adsabs.harvard.edu/abs/2022JATIS...8a1024L}
}

@ARTICLE{1968ApJ...153....1P,
       author = {{Peebles}, P.~J.~E.},
        title = "{Recombination of the Primeval Plasma}",
      journal = apj,
         year = 1968,
        month = jul,
       volume = {153},
        pages = {1},
          doi = {10.1086/149628},
       adsurl = {https://ui.adsabs.harvard.edu/abs/1968ApJ...153....1P}
}

@ARTICLE{1968ZhETF..55..278Z,
       author = {{Zeldovich}, Y.~B. and {Kurt}, V.~G. and {Syunyaev}, R.~A.},
        title = "{Recombination of Hydrogen in the Hot Model of the Universe}",
      journal = {Zhurnal Eksperimentalnoi i Teoreticheskoi Fiziki},
         year = 1968,
        month = jul,
       volume = {55},
        pages = {278-286},
       adsurl = {https://ui.adsabs.harvard.edu/abs/1968ZhETF..55..278Z}
}

@ARTICLE{McQuinn:2012,
       author = {{McQuinn}, Matthew and {O'Leary}, Ryan M.},
        title = "{The Impact of the Supersonic Baryon-Dark Matter Velocity Difference on the $z \!\sim\! 20$ 21cm Background}",
      journal = apj,
         year = 2012,
        month = nov,
       volume = {760},
       number = {1},
          eid = {3},
        pages = {3},
          doi = {10.1088/0004-637X/760/1/3},
archivePrefix = {arXiv},
       eprint = {1204.1345},
 primaryClass = {astro-ph.CO},
       adsurl = {https://ui.adsabs.harvard.edu/abs/2012ApJ...760....3M}
}

@ARTICLE{OLeary:2012,
       author = {{O'Leary}, Ryan M. and {McQuinn}, Matthew},
        title = "{The Formation of the First Cosmic Structures and the Physics of the $z \!\sim\! 20$ Universe}",
      journal = apj,
         year = 2012,
        month = nov,
       volume = {760},
       number = {1},
          eid = {4},
        pages = {4},
          doi = {10.1088/0004-637X/760/1/4},
archivePrefix = {arXiv},
       eprint = {1204.1344},
 primaryClass = {astro-ph.CO},
       adsurl = {https://ui.adsabs.harvard.edu/abs/2012ApJ...760....4O}
}

@article{Conroy_2008,
   title={Thermal Balance in the Intracluster Medium: Is {AGN} Feedback Necessary?},
   volume={681},
   ISSN={1538-4357},
   url={http://dx.doi.org/10.1086/587861},
   DOI={10.1086/587861},
   number={1},
   journal={The Astrophysical Journal},
   publisher={American Astronomical Society},
   author={Conroy, Charlie and Ostriker, Jeremiah P.},
   year={2008},
   month=jul, pages={151–166} }

@article{Kim_2005,
   title={Dynamical Friction and Cooling Flows in Galaxy Clusters},
   volume={632},
   ISSN={1538-4357},
   url={http://dx.doi.org/10.1086/432976},
   DOI={10.1086/432976},
   number={1},
   journal={The Astrophysical Journal},
   publisher={American Astronomical Society},
   author={Kim, Woong‐Tae and El‐Zant, Amr A. and Kamionkowski, Marc},
   year={2005},
   month=oct, pages={157–168} }

@article{Suzuguchi:2024btk,
    author = "Suzuguchi, Tomoya and Sugimura, Kazuyuki and Hosokawa, Takashi and Matsumoto, Tomoaki",
    title = "{Gas Dynamical Friction on Accreting Objects}",
    eprint = "2401.13032",
    archivePrefix = "arXiv",
    primaryClass = "astro-ph.GA",
    doi = "10.3847/1538-4357/ad34af",
    journal = "Astrophys. J.",
    volume = "966",
    number = "1",
    pages = "7",
    year = "2024"
}

@article{Kim_2009,
   title={NONLINEAR DYNAMICAL FRICTION IN A GASEOUS MEDIUM},
   volume={703},
   ISSN={1538-4357},
   url={http://dx.doi.org/10.1088/0004-637X/703/2/1278},
   DOI={10.1088/0004-637x/703/2/1278},
   number={2},
   journal={The Astrophysical Journal},
   publisher={American Astronomical Society},
   author={Kim, Hyosun and Kim, Woong-Tae},
   year={2009},
   month=sep, pages={1278–1293} }

@article{Hardy:2016mns,
    author = "Hardy, Edward",
    title = "{Miniclusters in the Axiverse}",
    eprint = "1609.00208",
    archivePrefix = "arXiv",
    primaryClass = "hep-ph",
    doi = "10.1007/JHEP02(2017)046",
    journal = "JHEP",
    volume = "02",
    pages = "046",
    year = "2017"
}

@article{Fukunaga:2020mvq,
    author = "Fukunaga, Hayato and Kitajima, Naoya and Urakawa, Yuko",
    title = "{Can axion clumps be formed in a pre-inflationary scenario?}",
    eprint = "2004.08929",
    archivePrefix = "arXiv",
    primaryClass = "astro-ph.CO",
    doi = "10.1088/1475-7516/2021/02/015",
    journal = "JCAP",
    volume = "02",
    pages = "015",
    year = "2021"
}

@ARTICLE{Just:1986,
       author = {{Just}, A. and {Kegel}, W.~H. and {Deiss}, B.~M.},
        title = "{Dynamical friction between the ISM and the system of stars}",
      journal = aap,
         year = 1986,
        month = aug,
       volume = {164},
       number = {2},
        pages = {337-341},
       adsurl = {https://ui.adsabs.harvard.edu/abs/1986A&A...164..337J}
}

@article{Ramirez:2022mys,
    author = "Ramirez, Edward D. and Buckley, Matthew R.",
    title = "{Constraining dark matter substructure with Gaia wide binaries}",
    eprint = "2209.08100",
    archivePrefix = "arXiv",
    primaryClass = "hep-ph",
    doi = "10.1093/mnras/stad2583",
    journal = "Mon. Not. Roy. Astron. Soc.",
    volume = "525",
    number = "4",
    pages = "5813--5830",
    year = "2023"
}

@article{Zhou:2021tvp,
    author = "Zhou, Huan and Lian, Yujie and Li, Zhengxiang and Cao, Shuo and Huang, Zhiqi",
    title = "{Constraints on the abundance of supermassive primordial black holes from lensing of compact radio sources}",
    eprint = "2106.11705",
    archivePrefix = "arXiv",
    primaryClass = "astro-ph.CO",
    doi = "10.1093/mnras/stac915",
    journal = "Mon. Not. Roy. Astron. Soc.",
    volume = "513",
    number = "3",
    pages = "3627--3633",
    year = "2022"
}

@article{Oguri:2017ock,
    author = "Oguri, Masamune and Diego, Jose M. and Kaiser, Nick and Kelly, Patrick L. and Broadhurst, Tom",
    title = "{Understanding caustic crossings in giant arcs: characteristic scales, event rates, and constraints on compact dark matter}",
    eprint = "1710.00148",
    archivePrefix = "arXiv",
    primaryClass = "astro-ph.CO",
    doi = "10.1103/PhysRevD.97.023518",
    journal = "Phys. Rev. D",
    volume = "97",
    number = "2",
    pages = "023518",
    year = "2018"
}

@article{Harigaya:2025pox,
    author = "Harigaya, Keisuke and Hu, Wayne and Liu, Rayne and Xiao, Huangyu",
    title = "{Universal lower bound on the axion decay constant from free streaming effects}",
    eprint = "2507.01956",
    archivePrefix = "arXiv",
    primaryClass = "astro-ph.CO",
    reportNumber = "FERMILAB-PUB-25-0430-T",
    doi = "10.1103/xgf4-xqjh",
    journal = "Phys. Rev. D",
    volume = "112",
    number = "6",
    pages = "063554",
    year = "2025"
}

@article{Murgia:2019duy,
    author = "Murgia, Riccardo and Scelfo, Giulio and Viel, Matteo and Raccanelli, Alvise",
    title = "{Lyman-{\ensuremath{\alpha}} Forest Constraints on Primordial Black Holes as Dark Matter}",
    eprint = "1903.10509",
    archivePrefix = "arXiv",
    primaryClass = "astro-ph.CO",
    reportNumber = "CERN-TH-2019-029",
    doi = "10.1103/PhysRevLett.123.071102",
    journal = "Phys. Rev. Lett.",
    volume = "123",
    number = "7",
    pages = "071102",
    year = "2019"
}

\clearpage
\onecolumngrid
\appendix
\begin{center}
\textbf{\Large
Supplemental Materials 
}
\end{center}

\section{\large Dynamical Friction in Gas}
\label{sec:I}

Dynamical friction (DF) refers to the deceleration experienced by a massive object moving through a background medium of lighter particles. In a collisional medium, such as baryons, the motion of the massive object gravitationally perturbs the surrounding gas, inducing an overdense region known as a wake. The gravitational attraction between the wake and the moving object exerts a drag force that slows down the object. Following Ref.~\cite{Ostriker_1999}, we can define the DF force of a perturber of mass $M$ moving through the baryon fluid as
\begin{equation}
    F_{\rm DF} = \frac{4 \pi G^2 M^2 \rho_{\rm B}}{v_{\rm rel}^2}\mathcal{I}(\mathcal{M},\Lambda),
    \label{eq:DFG}
\end{equation}
where  $v_{\rm rel}$ is the relative velocity of the perturber with respect to the rest frame of the gas, and $\rho_{\rm B}$ is the mass density of the gas. 
$\mathcal{I}$ is the result of integrating the gravitational force decelerating the wake over the entire wake, and is a function of the Mach number $\mathcal{M} \equiv v_{\rm rel} / c_s$ with $c_s$ being the sound speed of the baryons, and the Coulomb logarithm $\log \Lambda$, defined simply as 
\begin{alignat}{1}
    \ln \Lambda \equiv \ln \left( \frac{b_{\max}}{b_{\min}} \right)\,.
    \label{eq:SM_Coulomb_log}
\end{alignat}
This function introduces two regulators that regulate the gravitational force between the perturber and the gas at small and large impact parameters $b_{\min}$ and $b_{\max}$ respectively, where the approximation of an isolated, point-like perturber breaks down. 
While there is some arbitrariness to any choice of these impact parameters, our results are only logarithmically sensitive to them. 

Under the assumption that the density perturbations generated by the perturber are small, and that the equations of motion can be expanded to linear order, the full expression for $\mathcal{I}(\mathcal{M},\Lambda)$ is
\begin{equation}
\mathcal{I}(\mathcal{M}, \Lambda) =
\begin{cases}
\dfrac{1}{2}\ln\!\left(\dfrac{1 + \mathcal{M}}{1 - \mathcal{M}}\right) - \mathcal{M},
&
\mathcal{M} < 1 - x_{\min},
\\[8pt]
\dfrac{1}{4}x_{\min} - \dfrac{\mathcal{M}}{2}
- \dfrac{1 - \mathcal{M}^{2}}{4x_{\min}}
+ \dfrac{1}{2}\ln\!\left(\dfrac{1 + \mathcal{M}}{x_{\min}}\right),
&
1 - x_{\min} < \mathcal{M} < \sqrt{1+x_{\min}^{2}},
\\[8pt]
-\dfrac{1}{4x_{\min}}
+ \dfrac{(\mathcal{M}-x_{\min})^{2}}{4x_{\min}}
+ \dfrac{1}{2}\ln\!\left(\dfrac{1 + \mathcal{M}}{x_{\min}}\right),
&
\sqrt{1+x_{\min}^{2}} < \mathcal{M} < 1 + x_{\min},
\\[8pt]
\dfrac{1}{2}\ln\!\left(\dfrac{\mathcal{M}+1}{\mathcal{M}-1}\right)
+ \ln\!\left(\dfrac{\mathcal{M}-1}{x_{\min}}\right),
&
\mathcal{M} > 1 + x_{\min}\,,
\end{cases}
\label{eq:I_def}
\end{equation}

where $x_{\rm min} \equiv (1+\mathcal{M})/\Lambda$. 
Note that in Ref.~\cite{Ostriker_1999}, only the $\mathcal{M} \ll 1$ and $\mathcal{M} \gg 1$ limits were obtained; however, both of these expressions diverge as $M \to 1$, which is slightly unsatisfactory, although in practice this is usually not a problem since typically $\Lambda \gg 1$, and the range of $\mathcal{M}$ for which $1 + x_{\min} < \mathcal{M} < 1 - x_{\min}$ is very small. 
By carefully carrying out the integration over the wake, one recovers the expressions above, which is continuous for all $\mathcal{M}$. 

In the subsonic limit with $\mathcal{M} \ll 1$, $\mathcal{I}(\mathcal{M},\Lambda) \approx \mathcal{M}^3/3$, and the DF force is suppressed. 
In the linear theory approximation, this can be understood from the fact that the gas density perturbations generated by the perturber is almost spherically symmetric.  
However, in the supersonic case, the perturbed density distribution lags behind the perturber in the shape of a cone. 
This asymmetrical distribution of gas around the perturber generates a significant drag force, peaking at $\mathcal{M} = 1$. When the motion is supersonic, the force of dynamical friction, and consequently the heating rate, depends on the Coulomb logarithm; thus, $b_{\min}$ and $b_{\max}$ must be chosen carefully in Eq.~\eqref{eq:SM_Coulomb_log}.
Our choices are detailed in the main body of the paper, but to facilitate the discussion here, we will restate them here. 
For MACHOs, we set $b_{\min}$ to be the Bondi–Hoyle–Lyttleton radius, \textit{i.e.} 
\begin{align}
     b_{\rm min, \, M} = \frac{GM}{c_s^2+v_{\rm rel}^2}\,,
\end{align}
motivated by the fact that gas within this radius get ultimately accretes onto the object, leading to no net DF momentum transfer~\cite{Suzuguchi:2024btk}. 

For extended objects, the minimum impact parameter is given by $0.35 \mathcal{M}^{0.6}$ times the characteristic scale of the object, following Ref.~\cite{Kim_2009}, which found that this choice led to good agreement of their simulations with the linear result in the subsonic regime. For axion minihalos, we set this scale to be the virial radius, \textit{i.e.}
\begin{align}
    b_{\rm min, \,a} & = 0.35\mathcal{M}^{0.6}r_{\rm vir} \,.
\end{align}
Note that our choice of $b_{\rm min, \,M}$ is appropriate for MACHOs that are smaller than their Bondi-Hoyle-Lyttleton radius: for other compact objects with some known finite extent, it would be better to make the same choice as with axion minihalos instead.  
The maximum impact parameter is $b_{\rm max} = \mathcal{M} R_J$~\cite{OLeary:2012}, where $R_J$ is the Jean's length given by $R_J = c_S \sqrt{\pi/G\, \rho_m} \sim c_s H^{-1}$, with $H$ being the Hubble parameter.
$b_{\rm max}$ is roughly the distance traveled by the perturber within the lifetime of the Universe, and so is an appropriate estimate of the size of the wake behind the perturber. 
This choice is also consistent with what was used in the simulations presented in Refs.~\cite{McQuinn:2012,OLeary:2012}.

Note that Eq.~\eqref{eq:DFG}, with the definition of $\mathcal{I}$ in Eq.~\eqref{eq:I_def}, were obtained by assuming that density perturbations caused by the perturber are small, and by expanding the fluid equations up to linear order in density. 
However, the solution for the density perturbations obtained under this assumption is not self-consistent, as the perturbations exceed 1 near the Mach cone for supersonic perturbers. 
In fact, the density perturbations always diverge on the cone itself. 
Simulations are therefore necessary to understand if the linear approximation is suitable. 
Ref.~\cite{Kim_2009} conducted such simulations, and made detailed comparisons between simulation results and the linear approximation. 
They found that in the supersonic regime, the DF force only depended on the quantity $\eta \equiv \mathcal{A} / (\mathcal{M}^2 - 1)$, where $\mathcal{A} \equiv GM/(c_s^2 r_s)$, and $r_s$ is the size of the perturber (taken to be a Plummer sphere in their simulations). 
For $\eta \leq 2$, the DF force in the linear approximation $F_{\rm lin}$ was found to be in good agreement with the force obtained in the simulation, $F_{\rm sim}$, while for $2 < \eta \leq 100$, they found that the ratio $F_{\rm sim} / F_{\rm lin}$ was well-approximated by a power law, 
\begin{alignat}{1}
    \frac{F_{\rm sim}}{F_{\rm lin}} = \left( \frac{\eta}{2} \right)^{-0.45} \,.
\end{alignat}

For MACHOs, with $r_s = b_{\rm min,\,M}$, we have $\mathcal{A} = \mathcal{M}^2 + 1$, and
\begin{alignat}{1}
    \eta_{\rm M} = \frac{\mathcal{M}^2 + 1}{\mathcal{M}^2 - 1} \,.
\end{alignat}
We therefore only expect the difference between $F_{\rm sim}$ and $F_{\rm lin}$ to exceed 50\% only when the Mach number lies between $1 < \mathcal{M} \leq 1.1$, corresponding to a very narrow range of redshifts, which we judge to be an acceptable approximation.

For axion minihalos, we have a distribution of halo masses $M_h$ as a function of $M_{h,c}$, with $\mathcal{A} = G M_h/(c^2 r_{\rm vir})$. 
We find numerically that for $M_h = \SI{e6}{\MSolar}$, the ratio $F_{\rm sim} / F_{\rm lin}$ can be between 0.25--1 over a range of redshifts $\Delta z / z \sim 1$ around the transition from supersonic to subsonic. 
Although this is a somewhat large ratio, we compensate for this approximately by only considering the DF force to be acting on halos with a mass below \SI{e6}{\MSolar}. 
Furthermore, by overestimating the drag force, we actually underestimate the heating effect, since such massive halos transition from supersonic to subsonic earlier, typically toward an epoch where baryons were in stronger thermal contact with the CMB, leading to less efficient heating.

We expect that a proper simulation of a population of halos which are constantly decelerating through a baryonic fluid with a time-varying sound speed will produce significantly different results compared to those reported in Ref.~\cite{Kim_2009}. 
In addition, the deceleration of these objects will significantly alter how baryons fall into potential wells and the formation of the first structures, which again must be simulated. 
We leave a proper determination of these effects from simulations to future work. 

\end{document}